\documentclass[prb,10pt,aps,superscriptaddress,amsmath,amssymb,twocolumn]{revtex4-2}
\usepackage{hyperref}
\usepackage[capitalize]{cleveref}
\hypersetup{
    colorlinks=true,
    linkcolor=[rgb]{.1,.1,.6},
    citecolor=[rgb]{.1,.1,.6},
    filecolor=[rgb]{.1,.1,.6},      
    urlcolor=[rgb]{.1,.1,.6},
}
\usepackage {textcomp}
\usepackage{graphicx}
\usepackage{dcolumn}
\usepackage{bm}
\usepackage{amsmath}
\usepackage{amssymb}
\usepackage[section]{placeins}

\begin{document}

\title{Decoherence induced by a sparse bath of two-level fluctuators: peculiar features of $1/f$ noise in high-quality qubits.}

\author{M.~Mehmandoost}
\email{m.mehmandoost@tudelft.nl}
\author{V.V.~Dobrovitski}
\email{Corresponding author, v.v.dobrovitski@tudelft.nl}

\affiliation{QuTech and Kavli Institute of Nanoscience, Delft University of Technology, Lorentzweg 1, 2628 CJ Delft, The Netherlands}

\date{\today}

\begin{abstract}
Progress in fabrication of semiconductor and superconductor qubits has greatly diminished the number of decohering defects, thus decreasing the devastating low-frequency $1/f$ noise and extending the qubits' coherence times (dephasing time $T_2^*$ and the echo decay time $T_2$). However, large qubit-to-qubit variation of the coherence properties remains a problem, making it difficult to produce a large-scale register where all qubits have a uniformly high quality.
In this work we show that large variability is a characteristic feature of a qubit dephased by a sparse bath made of many ($n\gg 1$) decohering defects, coupled to the qubit with similar strength. 
We model the defects as two-level fluctuators (TLFs) whose transition rates $\gamma$ are sampled from a log-uniform distribution over an interval $[\gamma_{m},\gamma_M]$, which is a standard model for $1/f$ noise. 
We investigate decoherence by such a bath in the limit of high-quality qubit, i.e.\ when the TLF density $d$ is small (the limit of sparse bath, with $d=n/w\ll 1$, where $n$ is the number of TLFs and $w=\ln{[\gamma_M/\gamma_{m}]}$ is the log-width of the distribution).
We show that different realizations of the bath produce very similar noise power spectra $S(f)\sim 1/f$, but lead to drastically different coherence times $T_2^*$ and $T_2$. Thus, 
the spectral density $S(f)$ does not determine coherence of a qubit coupled to a sparse TLF bath, as opposed to a dense bath; instead,
decoherence is controlled by only a few exceptional fluctuators, determined by their value of $\gamma$. We show that removing only two of these TLFs greatly increases $T_2$ and $T_2^*$ times.
%
Our findings help theoretical understanding and further improvements in the coherence properties of semiconductor and superconductor qubits, battling the $1/f$ noise in these platforms.
\end{abstract}

\maketitle

\section*{Introduction}
Coherence times of many solid-state qubits are limited by the noise with spectral density
$
S(f) \propto 1/f,
$
where $f$ is the linear frequency. For instance, coherence times of spin qubits in semiconductor quantum dots are shortened by the background electric field noise with the $1/f$ spectral density \cite{Zwanenburg2013,VandersypenPhysicsToday2019,BurkardLaddPanNicholReviewQuDots23,Vandersypen2017,Kawakami2016,Yoneda2018,yoneda_noise-correlation_2023,Connors2022,Connors2020,Struck2020,rojas-arias_spatial_2023,SpenceNiquetMeunierEtalChargeNoise22,Petit2018,Petit2020,Stano2022,elsayed2022low,ChanMorelloDzurakPRApp18}. 
Similarly, coherence times of various superconducting qubits are often limited by the Josephson critical current noise and/or the magnetic flux noise, both having $1/f$ spectral density \cite{Ithier2005,Yoshihara2006,Martinis2005,Bialczak2007, Kakuyanagi2007,QuintanaEtal1fnoisePRL17,deGraaf2017,Luthi2018,Mueller2019,Lisenfeld2019,Schloer2019,Klimov2018,Meissner2018,paladino2014rmp,Lee2014,Wang2015,Rower2023evolution,Kumar2016}.

This noise is often produced by ensembles of two-level fluctuators (TLFs) \cite{Kogan1996,DuttaHorn1981rmp,Anderson1972,Phillips1972,McWhorter1957}, i.e.\ two-state systems undergoing random transitions between the two states at the rate $\gamma$.
In quantum dot devices, the TLFs are often associated with charge traps, which randomly trap/release electrons, or where the trapped charge can randomly jump between two nearby positions via tunneling or thermal activation, 
such as the two-level systems in dielectric layers and in the vicinity of semiconducting layers.
The randomly fluctuating electric fields, produced by the charge traps, distort and displace the orbital wavefunction of the electron confined in the quantum dot, and these distortions affect the electron spin due to, for instance, spatial inhomogeneity of the $g$-factor of the spatially inhomogeneous magnetic field created by a micromagnet \cite{PioroLadriere2008,Kawakami2016,ShalakDelerueNiquetChargeNoiseSiHole23,ShehataVanDorpeEtalChargeNoiseQuDots23,yoneda_noise-correlation_2023,PaqueletWuetzScapucciRussEtalChareNoiseSi23,Yoneda2018,Connors2020,Connors2022,Struck2020,rojas-arias_spatial_2023,SpenceNiquetMeunierEtalChargeNoise22,Meyer2023electrical,KepaCywinskiKrzywdaSpinNoise23,KepaFockeCywinskiKrzywdaChargeNoise23,Ahn2021}. 
In superconducting devices, the TLFs located in the insulating barriers of Josephson junctions may randomly modulate the Josephson critical current; another likely candidate for the TLFs are the magnetic impurities producing the magnetic flux noise \cite{paladino2014rmp,deGraaf2017,Mueller2019,Lisenfeld2019,Paladino2002,Lee2014,Wang2015,Faoro2006,Faoro2008,Szankowski2017,Luthi2018,Mueller2019,Klimov2018,Meissner2018,deGraaf2017,Schloer2019,Wang2015,Lisenfeld2023enhancing,Rower2023evolution,Kumar2016}. 

TLFs can affect both the qubit lifetime $T_1$ and the qubit dephasing times,  the free dephasing time $T_2^*$ (measured e.g.\ in the Ramsey decay experiments) and the echo decay time $T_2$.
In many superconducting qubits \cite{Martinis2005,Klimov2018,Lisenfeld2019,Lisenfeld2023enhancing,Schloer2019,Burnett2019} the former effect is dominant, and is caused by the resonant TLFs, i.e.\ by those defects, for which the energy difference between the two relevant defect states is close to the energy difference $E_0=\hbar\omega_0$ between the qubit's states $|0\rangle$ and $|1\rangle$. Such TLFs can undergo resonant flip-flop transitions with the qubit, 
making the qubit to switch between the states $|0\rangle$ and $|1\rangle$, thus reducing its lifetime $T_1$.
The effect of non-resonant TLFs, for which the energy difference between the two states is much smaller than $E_0$, is limited to pure dephasing.
The dephasing caused by a bath of non-resonant TLFs is considerable for many superconducting qubits, and is critical for majority of quantum dot-based spin qubits, including highly promising SiMOS and Si/SiGe platforms. This type of baths is the subject of our studies below.

The conventional approach to decoherence by an ensemble of TLFs relates the decoherence profile to the noise power spectrum $S(f)$, i.e.\ to the dynamics of the whole TLF ensemble, see e.g.\ reviews \cite{Szankowski2017,paladino2014rmp}. 
This approach uses the standard model of $1/f$ noise 
\cite{DuttaHorn1981rmp}, 
assuming a large number of TLFs coupled to the qubit with more or less comparable strengths, while the values of $\gamma$ for each TLF are drawn independently at random from a log-uniform distribution within the broad range $[\gamma_m,\gamma_M]$ ($\gamma_{M}\gg\gamma_{m}$); then the noise experienced by the qubit has $1/f$ spectral density for frequencies in the range $\gamma_{m}\lesssim 2\pi f\lesssim\gamma_{M}$, see Fig.~\ref{fig:Fig1}. Within the 
conventional approach, the decay of the qubit's coherence is treated in a central limit theorem-like manner, using the cumulant expansion and retaining only the second-order cumulant (Gaussian approximation); it predicts that (i) many TLFs in the bath contribute to decoherence of the qubit, and (ii) the qubits having the same noise spectrum $S(f)$ should exhibit the same coherence decay.

Recent progress in fabrication and in controlling the materials properties has greatly diminished the number of defects producing $1/f$ noise  \cite{Yoneda2018,Connors2022,Kumar2016,deGraaf2017,PaqueletWuetzScapucciRussEtalChareNoiseSi23,elsayed2022low,Meyer2023electrical,Lisenfeld2023enhancing,RyuKangDevitalizeDefects22}, thus 
significantly improving the qubit quality.
However, variability in the coherence times between different qubits (or of the same qubit observed over many hours/days) remains high \cite{Stano2022,Klimov2018,Schloer2019,Mueller2019,Martinez2022,Luthi2018,Connors2022}, making it difficult to create a chip with a many-qubit register where all qubits would be stable and have uniformly good coherence. 
Besides, there is a growing amount of theoretical and experimental evidence \cite{Schloer2019,Ahn2021,Connors2020,Connors2022} that in modern high-quality qubits the coherence decay is controlled by only a handful of TLFs (possibly, even a single TLF).
These features, being at odds with the 
conventional theories, prompt for a deeper theoretical analysis of the modern high-quality solid-state qubits.

In this work we study a model for a high-quality qubit decohered by $1/f$ noise
\footnote{Data and codes used in this work could be accessed at \url{https://zenodo.org/records/10988052}, DOI 10.5281/zenodo.10988052}.
We assume, in a standard manner, the noise to be created by an ensemble of TLFs, all coupled with comparable strength to the qubit, with log-uniform distribution of the transition rates $\gamma$. However, we focus on the non-conventional situation of sparse bath, when the number $n$ of TLFs coupled to the qubit is large, but the range $[\gamma_{m},\gamma_{M}]$ is even larger, such that the TLF density $d=n/w$ is small ($w=\ln{[\gamma_M/\gamma_{m}]}$ is the log-width of the distribution). We demonstrate, both analytically and numerically, how the conventional approach,
based on the central limit theorem, breaks down at small densities $d$, even for large number $n$ of TLFs. We also explain why the Gaussian approximation, employed in the conventional approach, requires not only the total number $n$, but also the density $d$, to be large. 
Namely, we show that for sparse baths, the qubit decoherence is controlled by only few exceptional TLFs, whose contribution dominates all other TLFs in the ensemble. These TLFs are determined not by exceptionally large coupling to the qubit, but by their transition rate $\gamma$. 
Removing or adding such TLFs to the bath greatly affects the dephasing times  $T_2$ and $T_2^*$, such that the coherence times fluctuate wildly, as the values of $\gamma$ of the exceptional TLFs randomly change from one realization of the bath (in one qubit) to another (in another qubit), although the noise spectral density $S(f)$ remains almost the same.

Non-Gaussian effects in the bath of TLFs, their possible origins, and their impact on the dynamics of qubit dephasing, have been explored and discussed before \cite{Galperin2006,Schriefl2006,Szankowski2017,Bergli2009,paladino2014rmp,Paladino2002} for other models and regimes, focusing mostly on the situation of broadly distributed coupling strengths. But, to our knowledge, the non-Gaussian behavior emerging solely from the low density of TLFs in sparse baths, even when many fluctuators are coupled with the same or similar strength to the qubit, has not been analyzed before,
and peculiarities of the sparse bath dynamics have not been studied yet; in fact, the role of the parameter $d$ or its analog has not been discussed in these works. In this respect, our work challenges and corrects some widely accepted statements. 

Emergence of exceptional TLFs in the sparse baths offers a possible way to bridge  the standard models, with many TLFs coupled to the qubit, with the recently proposed models \cite{Schloer2019,Ahn2021,Connors2020}, where only few TLFs are noticeably coupled to the qubit 
\footnote{Emergence of special TLFs, which affect the decoherence particularly strongly, was observed in a class of models where many TLFs are coupled to the qubit and the coupling constants vary in a broad range \cite{Galperin2006,Schriefl2006,Bergli2009}: in these models, the special TLFs are determined by the optimal coupling strength. 
But in the case of sparse many-TLF bath, the origin of exceptional TLFs is different: it is not the coupling strength, but the special value of $\gamma$, which becomes important for $d\ll 1$.}.
Also, the $1/f$-type noise spectrum is a natural feature of the many-TLF baths (both dense and sparse), while in the few-TLF model  \cite{Schloer2019,Ahn2021,Connors2020} it is the consequence of the coupling of the TLFs to their own environment (decoupled from the qubit but coupled to the TLFs). 

The rest of the paper is organized as follows. In Sec.~\ref{sec:two} we describe the model for a many-TLF bath coupled to the qubit, and demonstrate that the noise power spectra have a $1/f$ character, with little sample-to-sample variation for the whole range of $d$. In Sec.~\ref{sec:three} we present analytical and numerical results, showing that the Ramsey and echo decay curves, in contrast with the noise power spectra, exhibit large sample-to-sample variations for sparse baths. We explain this effect analytically, and show that the Ramsey decay is controlled by the properties of only a few exceptional TLFs, whose parameters vary from sample to sample. 
In Sec.~\ref{sec:excTLF} we demonstrate numerically that for sparse baths,  removal of only two exceptional TLFs makes decoherence substantially slower, for both individual baths and large ensembles of baths. We identify these exceptional TLFs for the Ramsey and for the echo decay. In Sec.~\ref{sec:improve} we demonstrate substantial improvements caused by the removal of only two exceptional TLFs, using the decay times and the qubit fidelities as metrics. In Sec.~\ref{sec:dispersv} we show that these results also hold for the baths with reasonably large spread in the coupling strengths of individual TLFs. In Sec.~\ref{sec:dd} we briefly discuss dynamical decoupling for the case of sparse baths, showing that significant increase in the number of pulses leads to a moderate slowdown of decoherence, but greatly decreases the sample-to-sample variability of the decay curves. Our findings are summarized and conclusions are presented in Sec.~\ref{sec:sum}.

\section{Model description}
\label{sec:two}

A large number of defects is unavoidably present in realistic super- and semiconducting qubits due to their meso- or macroscopic size. However, if the qubit is fabricated with utmost care then the most egregious defects, particularly strongly coupled to the qubit, are eliminated (or avoided) during fabrication or deactivated at the post-fabrication stage \cite{Meyer2023electrical,Lisenfeld2023enhancing,RyuKangDevitalizeDefects22}; the currently manufactured high-quality semiconducting and superconducting qubits seem to approach this limit. Thus, it is natural to assume that the remaining defects are weakly coupled to the qubit, with more or less similar strengths, although other parameters may differ substantially.
We represent each defect as a two-level system, incoherently tunneling \cite{McWhorter1957} or thermally hopping \cite{DuPre1950,VanDerZiel1950} between two relevant states; a broad range of defects, including the charge traps near the gate electrodes, magnetic impurities, etc., under various conditions, are adequately represented in this way \cite{Paladino2002, Weissman1993, Mueller2019,Galperin2006}.

The qubit is described as a pseudo-spin 1/2 in the usual way, using the corresponding Pauli matrices 
$\sigma_x=|0\rangle\langle 1|+|1\rangle\langle 0|$, 
${\sigma}_y=-i|0\rangle\langle 1|+i|1\rangle\langle 0|$, 
and ${\sigma}_z=|0\rangle\langle 0|-|1\rangle\langle 1|$, where $|0\rangle$ and $|1\rangle$ are the computational basis states of the qubit, separated in energy by $E_0=\hbar\omega_0$.
In the coordinate frame which rotates with the qubit's Larmor frequency $\omega_0$ (rotating frame \cite{slichter}), the Hamiltonian describing decoherence of the qubit by the bath of TLFs has the form 
\begin{equation}
\label{eq:Ham}
{H} = \frac{1}{2} \sigma_z\,B(t), \quad B(t)=\sum_{k=1}^n v_k \xi_k(t),
\end{equation}
where we took into account that the TLFs under consideration are non-resonant, and thus omitted the terms proportional to ${\sigma}_x$ and ${\sigma}_y$. The noise $B(t)$, acting on the qubit, 
is created by an ensemble of $n$ statistically independent TLFs. The dynamics of each TLF are described as a random telegraph process \cite{vanKampen2007,GardinerRandomProc}, that is the $k$-th fluctuator is represented by a two-state random stationary Markov process $\xi_k(t)$, which can assume two values $+1$ and $-1$, and makes random transitions between them with the rate $\gamma_k$, equal for transitions in both directions (symmetric fluctuator, typical for various defects in a wide range of conditions \cite{BlackHalperin77,Laikhtman85,paladino2014rmp}), such that the correlation function for the $k$-th TLF is
\begin{equation}
\langle\xi(t)\xi(t+s)\rangle = \exp(-2\gamma_k|s|).
\end{equation}
Representing quantum two-level defect by a classical random process is justified when the coupling strength $v_k$ of the TLF to the qubit is smaller than the decoherence rate of the TLF itself \cite{Wold2012}. 
Following the standard convention, here and below we set $\hbar=1$, expressing all energies and the coupling strengths in the same units as the rates and the angular frequencies (${\rm s}^{-1}$).

The properties of the TLF bath are characterized by the set 
\begin{equation}\label{eq:quenchedBathSet}
\mathcal{B}=\{(\gamma_1,~v_1),\dots,(\gamma_k,~v_k),\dots,(\gamma_n,~v_n)\},
\end{equation}
of the coupling constants and transition rates of all TLFs, $k=1,\dots n$. 
The rates $\gamma_k$ are independently sampled from the log-uniform distribution 
\begin{equation}\label{eq:PGamma}
P_\Gamma(\gamma)=\frac{1}{w\ \gamma},\quad \gamma\in[\gamma_{m},\gamma_{M}],
\end{equation}
with sharp lower and higher cutoffs $\gamma_{m}$ and $\gamma_{M}$, respectively; the normalization constant $w=\ln(\gamma_{M}/\gamma_{m})$ is the log-width of the distribution. 
Such a distribution is natural for many two-state defects. Whether the transitions between the states are caused by tunneling or thermal excitations, the corresponding rates have the form $\gamma=R\,{\rm e}^{-\lambda}$; here $R^{-1}$ is a characteristic internal timescale of the defect, and $\lambda$ is either a tunneling exponent, or, in the case of thermally induced hopping, the Arrhenius factor $\lambda=E_b/kT$, where $E_b$ is the height of the energy barrier separating the two states of the defect, $k$ is the Boltzmann constant, and $T$ is the temperature \cite{BlackHalperin77,Laikhtman85,paladino2014rmp}. For many relevant defects, the parameter $\lambda$ is distributed uniformly between the values $\lambda_{m}$ and  $\lambda_{M}$, such that the transition rate $\gamma\propto{\rm e}^{-\lambda}$ has the log-uniform probability density (\ref{eq:PGamma}).
The couplings $v_k$ are also taken as a set of independent identically distributed random variables. We assume them to be comparable for all TLFs, such that the distribution $P_V$ for each $v_k$ is taken either as the Dirac delta distribution (with all $v_k=\overline{v}$, $k=1,\dots n$), or the normal distribution with finite width $\sigma_v$. The opposite situation of a very broad distribution of the couplings has been considered before \cite{Galperin2006,Bergli2009,Schriefl2006,Paladino2002}.
%

As will be shown below, an important parameter determining the dynamics of decoherence is the density of the TLFs,
\begin{equation}
d=\frac{n}{w}.
\end{equation} 
We will refer to the case of small or large $d$ as a sparse or dense bath, respectively.

\begin{figure}[tbp!]
\centering
\includegraphics[width=1\linewidth]{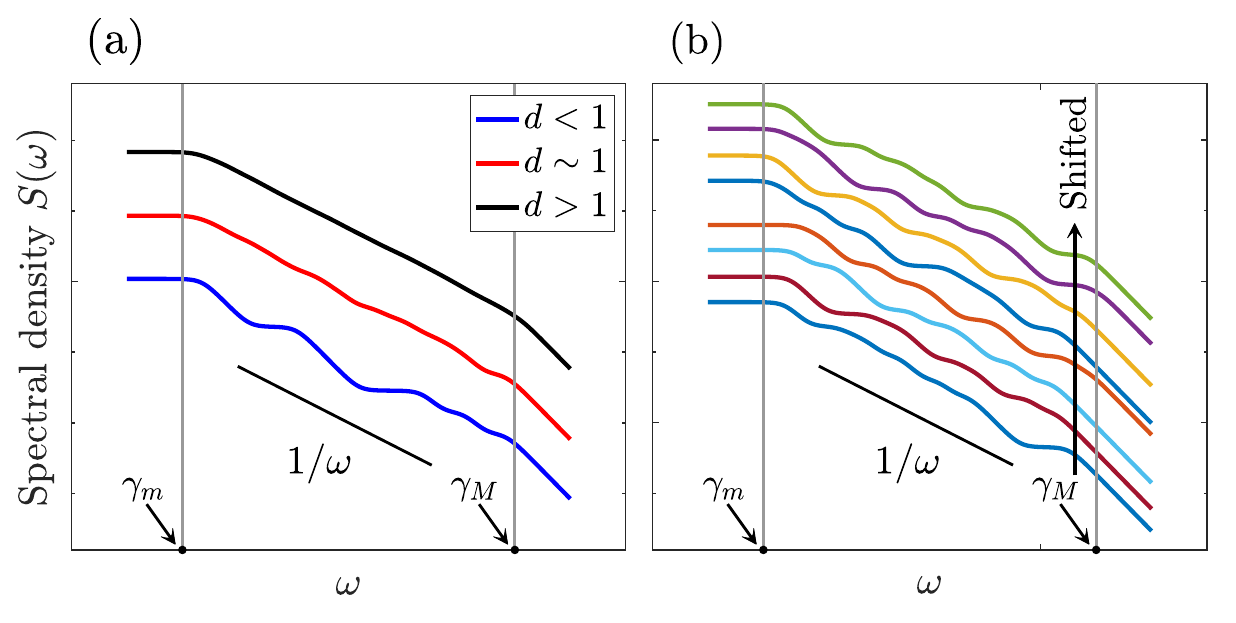}
\caption{Simulated spectral densities of the noise produced by bath samples of varying density $d$. Both plots (a) and (b) have the same ratio of $\gamma_M/\gamma_m=10^9$. The noise variance $n\bar{v^2}=\gamma_m^2$ is the same for all plotted spectra, but the spectra are shifted for clarity. The horizontal and vertical axes in both plots are scaled logarithmically (log-log plots). {\bf (a)} Spectral densities of sample baths with different values of $d$. The number of TLFs for the blue, red, and black curves are respectively $n=5,~30,~1000$, resulting in bath densities of $d\approx0.18,~1.08,~36.19$. {\bf (b)} Spectral densities of sparse sample baths with fixed values of $n$ and $d$. For all spectra $n=10$, such that $d\approx0.36$.}
\label{fig:Fig1}
\end{figure}

For a given qubit, the set $\mathcal{B}$ of the bath parameters is time-independent, meaning that the bath is \textit{quenched}, such that the random process $B(t)$ is stationary, with zero mean and the variance
\begin{equation}
\label{eq:Bvariance}
\beta(\mathcal{B})\equiv \langle B^2(t)\rangle =\sum_{k=1}^n v_k^2,
\end{equation} 
where the angular brackets $\langle\dots\rangle$ denote averaging over the noise realizations with a fixed set of parameters $\mathcal{B}$. The symbol $\mathcal{B}$ here and below reminds us that the properties of the noise $B(t)$ depend on the specific set $\mathcal{B}$, corresponding to the given realization of the quenched bath. 
The power spectrum (first spectral density) of the stationary noise $B(t)$ is given by the Fourier transform of its two-point correlation function \cite{Kubo1991,vanKampen2007,GardinerRandomProc}, i.e.,
\begin{equation}\label{eq:SpecDensDef}
S(\mathcal{B};\omega) = \int_{-\infty}^{+\infty}\,d\tau\,\mathrm{e}^{i\omega s} \langle B(t) B(t+s)\rangle,
\end{equation}
where $\omega = 2\pi f$. For the TLF bath in question, the power spectrum is a sum 
\begin{equation}\label{eq:Lorentzian}
S(\mathcal{B};\omega)=\sum_{k=1}^n \frac{4\gamma_k v_k^2}{4\gamma_k^2+\omega^2},
\end{equation}
consisting of the Lorentzian contributions from each independent random telegraph process $\xi_k(t)$, whose correlation function is 
$\langle\xi_k(t)\xi_k(t+s)\rangle=\exp{(-2\gamma_k|s|)}$. 

Several representative examples of the power spectra $S(\mathcal{B};\omega)$ for such baths are shown in Fig.~\ref{fig:Fig1}. Each spectrum was obtained via direct summation of Lorentzians in Eq.~\ref{eq:Lorentzian}, with the TLF parameters sampled from the distributions $P_\Gamma(\gamma)$ given by Eq.~\ref{eq:PGamma}, assuming $v_k=\overline{v}$ for all $k$.

We remind that the presented spectra correspond to the quenched baths, and do not involve averaging over the bath realizations.
While each spectrum is a sum of Lorentzians, the overall sum, because of the log-uniform distribution of the rates $\gamma_k$, has $1/f$ dependence for both sparse and dense baths. Even for sparse baths with $d\approx 0.36$, the spectra $S(\mathcal{B};\omega)$ remain practically the same for different bath samples, and retain clear $1/f$ character, see Fig.~\ref{fig:Fig1}b. Little wiggles, produced by the individual Lorentzians corresponding to individual TLFs, are barely visible, and their positions vary between different realizations of the bath, and already for $d\sim 1$ such small variations are practically invisible (Fig.~\ref{fig:Fig1}a).

Our simulations confirm that for both sparse and dense baths, the sample-to-sample variations in the noise spectrum remain small, and all spectra remain close to the average noise spectrum
\begin{equation}
{\overline{S}}(\omega)=\int\,S(\mathcal{B};\omega)\,\prod_k P_V(v_k) dv_k\,P_\Gamma(\gamma_k) d\gamma_k,
\end{equation}
with $S(\mathcal{B};\omega)$ given by Eq.~\ref{eq:Lorentzian}. Since $v_k$ and $\gamma_k$ are independent, and each of them is identically distributed with the corresponding distribution $P_V(v_k)$ or $P_\Gamma(\gamma_k)$, the average spectrum is easily calculated:
\begin{equation}
{\overline{S}}(\omega)=\frac{2n \overline{v^2}}{w\cdot\omega} 
\left[\arctan{\left(\frac{2\gamma_{M}}{\omega}\right)} - \arctan{\left(\frac{2\gamma_{m}}{\omega}\right)}\right],
\end{equation}
where $\overline{v^2}=\int v^2\,P_V(v)\,dv$. This spectral density has the form $1/\omega$ for $\gamma_{m}\ll\omega\ll\gamma_{M}$, with the $1/\omega^2$ cutoff at high frequencies $\omega\gg\gamma_{M}$ and flat cutoff at $\omega\ll\gamma_{m}$.

These results are in agreement with the conventional wisdom about good self-averaging properties of the $1/f$ noise spectrum between $\gamma_m$ and $\gamma_M$ \cite{paladino2014rmp,Galperin2006,Bergli2009}, and confirm that this remains true even for sparse TLF baths. However, our results on the qubit dephasing under the influence of the sparse baths, presented below in Sec.~\ref{sec:three}, show a very different picture.

\section{Dephasing dynamics of a qubit coupled to a sparse bath}
\label{sec:three}

The noise $B(t)$ in the Hamiltonian (\ref{eq:Ham}) does not change the populations of the qubit states $|0\rangle$ and $|1\rangle$, but affects the phase between these states, leading to dephasing of the qubit. The rate of dephasing can be assessed in experiments by measuring the decay of the Ramsey signal $F(t)$ with time. Initially, the qubit is prepared in the state $(|0\rangle+|1\rangle)/\sqrt{2}$; using the standard parametrization of the qubit density matrix as 
$
{\rho} = (1/2)\left[{\mathbf 1} + m_x {\sigma}_x + m_y {\sigma}_y + m_z {\sigma}_z\right]
$,
where ${\mathbf 1}$ is a 2$\times$2 identity matrix, the initial state corresponds to $m_x=1$ and $m_y=m_z=0$. Then the qubit is left to evolve freely for the time $t$ under the action of the noise $B(t)$, such that the value of $m_x(t)$ is given by the real part of the noise-averaged exponent
\begin{equation}
m_x(t)\equiv F(\mathcal{B};t)= 
{\rm Re}\ \left\langle\exp{\left(-i\int_0^t B(s)\,ds\right)}\right\rangle,
\end{equation}
while the value of $m_z$ remains zero; this is also true for $m_y$ because the statistics of the noise $B(t)$ remains invariant under the change  
$B(t)\to -B(t)$. Since the noise is a sum of $n$ independent telegraph processes 
$\xi_k(t)$, the Ramsey decay function $F(\mathcal{B};t)$ is a product of individual contributions $f_k(\gamma_k,v_k;t)$ from each TLF:
\begin{equation}
\label{eq:RamseyHahnProduct}
F(\mathcal{B};t)= \prod_{k=1}^n f_k(t),
\end{equation} 
where the individual terms are
\begin{equation}
f_k(t)\equiv f(\gamma_k,v_k;t)=\left\langle\exp{\Bigl(-i\,v_k\int_0^t \xi_k(s)\,ds\Bigr)}\right\rangle.
\end{equation}
Here we used the notation $F(\mathcal{B};t)$ to emphasize that the Ramsey decay function depends on the specific bath parameter set $\mathcal{B}$; in the text below we sometimes omit this variable if the relevant parameter set $\mathcal{B}$ is known from the context or is not relevant. 

An explicit expression for the Ramsey decay function $f(\gamma,v;t)$ for an individual TLF is known \cite{Paladino2002,Zhidomirov1969,deSousa2003,Galperin2006,RamonNonGauss2015,Mkhitaryan2014},
\begin{equation}
\label{eq:RamseySingleTLF}
f(\gamma,v;t) = {\rm e}^{-\gamma t}(\cosh{\alpha t} + \frac{\gamma}{\alpha}\sinh{\alpha t}), 
\end{equation}
where $\alpha=(\gamma^2 - v^2)^{1/2}$.
%
%
The overall shape of the decay function $f(\gamma,v;t)$ is determined by the relation between $v$ and $\gamma$.

For $v\gg\gamma$, the decay happens mostly at $t\lesssim\gamma^{-1}$, and if the qubit is coupled to $n\gg 1$ such TLFs, then the overall decay happens very quickly, on a timescale of the order of $t_{s}=(v\sqrt{n})^{-1}$. Indeed, when  $v\gg\gamma$, at short times $f(\gamma,v;t)\approx \exp{(-v^2 t^2/2)}$, and the product of $n\gg 1$ such factors leads to the decay $F(t)=\exp{(-n v^2 t^2/2)}$, the standard result for quasi-static Gaussian dephasing.
In the opposite case, $v\ll\gamma$, the decay happens almost completely at 
$t\gtrsim \gamma^{-1}$, when the decay shape is practically exponential, corresponding to the regime of motional narrowing \cite{slichter,Kubo1991}, $f(\gamma,v;t)\approx \exp{[-v^2 t \gamma^{-1}/2]}$. For our work, the latter case is of most physical interest and importance, and we focus on the situation when most TLFs are motionally narrowed, with $v_k\ll\gamma_k$.

\begin{figure}[tbp!]
\centering
\includegraphics[width=1\linewidth]{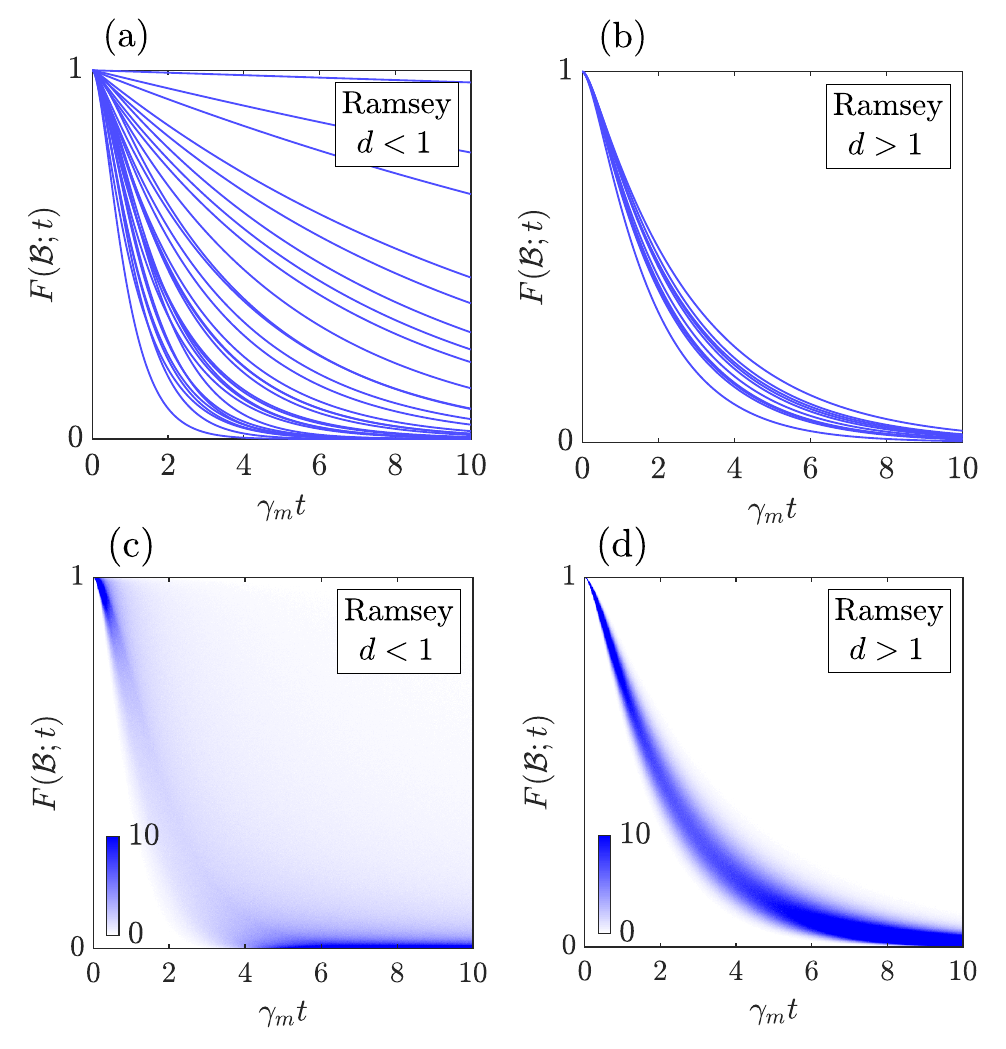}
\caption{Ramsey decay for sparse versus dense baths. The ratio $\gamma_M/\gamma_m=10^5$ is fixed in all panels. {\bf (a)} Samples of the Ramsey decay curves $F(t)$
for sparse baths with $n=10$ TLFs, which corresponds to $d\approx0.87$; the coupling strengths are the same for all TLFs, $v_k=\bar{v}=\gamma_m$. Thirty randomly chosen samples are shown. {\bf (b)} Samples of the Ramsey decay $F_R(t)$ 
for dense baths with $n=160$, which corresponds to $d\approx 13.9$; the coupling strengths are the same for all TLFs, $v_k=\bar{v}=0.25\gamma_m$. The value of $\bar{v}$ is reduced in comparison with panel (a), such that the variance $\beta(\mathcal{B})$ of the field $B(t)$ is the same in both panels, see text for details. Ten randomly chosen samples are shown; the number of curves is reduced in comparison with panel (a) to make the individual curves clearly visible. 
{\bf (c)},{\bf (d)}: Two-dimensional histograms of the estimated p.d.f.'s $P_F(F,t)$ of the Ramsey decay curves, presented in a qualitative form, for sparse baths, panel {\bf (c)}, and for dense baths, panel {\bf (d)}. Each panel comprises $10^9$ two-dimensional rectangular bins (pixels), obtained by dividing the time axis $t$ into $l_t=10^3$ bins, and the axis $F$ into $l_F=10^3$ bins. The value of $P_F(F,t)$ was estimated by drawing $M=10^5$ samples of the bath parameter sets $\mathcal{B}_m$ ($m=1,\dots M$) from the corresponding distributions, and calculating the Ramsey decay function $F_m(t)$ for each $\mathcal{B}_m$. Every $F_m(t)$ was evaluated at the center $t_p$ of every time bin ($p=1,\dots l_t$), and, according to the value $F_m(t_p)$, the point was added to the appropriate bin along the axis $F$. 
The total number $K(q,p)$ of the points in the pixel centered at $(F_q,t_p)$ is taken as an estimate for $P_F(F,t)$ at the point $(F=F_q,t=t_p)$; see the text for more details.
For the figures, the total number $K(q,p)$ in each pixel is divided by 100, and the values smaller than 10 are linearly mapped to the color bar, while those greater than or equal to 10 are mapped to the darkest shade of the color bar.} 
\label{fig:Fig2}
\end{figure}

Statistical properties of the dephasing curves for sparse and dense baths are compared in Fig.~\ref{fig:Fig2}. For these figures, a large number  
$M=10^5$ of independent samples of the bath parameter sets $\mathcal{B}$ was produced. The values $\gamma_k$ for each TLF ($k=1,\dots n$) for every set $\mathcal{B}_m$ ($m=1,\dots M$)  were independently drawn from the distribution $P_\Gamma(\gamma)$ given by Eq.~\ref{eq:PGamma}, with certain cutoff values $\gamma_{m}$ and $\gamma_{M}$, while the couplings $v_k$ were chosen equal, $v_k=\bar{v}$, for all TLFs.
The Ramsey decay curves $F_m(t)\equiv F(\mathcal{B}_m;t)$ were calculated for each bath sample $\mathcal{B}_m$ as a product (see Eq.~\ref{eq:RamseyHahnProduct}) of contributions $f_k(t)$ of individual TLFs given by Eq.~\ref{eq:RamseySingleTLF}. 
Several randomly chosen samples of the dephasing curves $F_m(t)$ for sparse and for dense baths are shown in Fig.~\ref{fig:Fig2}(a) and (b), respectively, with the corresponding densities $d\approx 0.87$ and $d\approx 13.90$.
In order to make a meaningful comparison between different baths with different number $n$ of TLFs, the value $\bar{v}$ was adjusted for each bath parameter set $\mathcal{B}$ in such a way that all baths have the same variance $\beta(\mathcal{B})$ of the noise $B(t)$ (see Eq.~\ref{eq:Bvariance}). In this manner we ensure that the noise spectral densities $S(\mathcal{B}_m;\omega)$  have the same overall amplitude  for all bath samples in the physically relevant region $\gamma_{m}\ll\omega\ll\gamma_{M}$.

In order to present cumulative statistics of the Ramsey decay curves $F_m(t)$ over all $M=10^5$ bath samples, in Fig.~\ref{fig:Fig2}(c) and (d) we show the two-dimensional density plots of the Ramsey decay curves. Namely, we divide the range $[0,1]$ of the quantity $F$ into $l_F=10^3$ bins, each of the length $\delta F=10^{-3}$; similarly, the range $[0,T_\mathrm{max}]$ of the time variable $t$ is divided into $l_t=10^3$ bins, each of the length $\delta t=T_\mathrm{max}\cdot 10^{-3}$. Thus, we obtain $10^9$ square bins (pixels) on the two-dimensional $F$--$t$ plane. For every function $F_m(t)$, we calculate a set of values $F_m(t_p)$ at the time points $t_p$ ($p=1,\dots l_t$), where $t_p$ is the center of the $p$-th time bin. The curve $F_m(t)$ is thusly represented as a set of $l_t$ points with the coordinates $(F=F_m(t_p),t=t_p)$ on the $F$--$t$ plane. Every such a point is placed in an appropriate bin along the axis $F$, centered at $F=F_q$; therefore, each curve fills $l_t$ pixels on the $F$--$t$ plane, one point per time bin. As more and more curves $F_m(t)$ are sampled, the number $K(q,p)$ of the points in the pixel centered at $(F_q,t_p)$ increases, and, with proper normalization, the value $K(q,p)$ gives an estimate for the probability density function (p.d.f.) $P_F(F,t)$ at the point $(F=F_q,t=t_p)$, thus characterizing the statistics of the curves $F_m(t)$ obtained for a given ensemble of the bath parameters. The resulting estimated p.d.f.'s are shown in Fig.~\ref{fig:Fig2}(c) and (d) in a qualitative form.

The stark contrast between the behavior of the dephasing curves in Fig.~\ref{fig:Fig2} and the noise spectra in Fig.~\ref{fig:Fig1} is obvious. For the dense baths, all dephasing curves $F_m(t)$ in Fig.~\ref{fig:Fig2}(b) stay close to each other, and the density $P_F(F,t)$ in Fig.~\ref{fig:Fig2}(d) is sharply concentrated around a single decay curve, such that the sample-to-sample variations in the Ramsey decay curves, and in the corresponding dephasing time $T_2^*$, are small. 
For the sparse baths, the dephasing curves in Fig.~\ref{fig:Fig2} (a) vary wildly between different bath samples, and the density $P_F(F,t)$ in Fig.~\ref{fig:Fig2} (c) is spread over large area, such that the sample-to-sample variations in the Ramsey decay signals and in the dephasing times $T_2^*$ are very large. This is drastically different from the behavior of the noise power density curves in Fig.~\ref{fig:Fig1}, which remain close to each for different realizations $\mathcal{B}_m$ of the bath parameters, for both sparse and dense baths. Thus, for sparse baths the relation between the Ramsey decay $F(\mathcal{B};t)$ and the noise spectrum $S(\mathcal{B};\omega)$, predicted by the conventional approach \cite{Cywinski2008prb,Schriefl2006,Szankowski2017,Galperin2006,Bergli2009,CywinskiWitzelDS2009}, does not hold. The central limit theorem-type treatment appears incorrect when applied to the sparse baths, even though the total number $n$ of TLFs is large.

In order to understand such a behavior at the qualitative level, let us consider the situation when all TLFs are motionally narrowed, with $\gamma_k\gtrsim \bar{v}$ for all $k=1,\dots n$. In that case, the Ramsey decay has the form (see Eqs.~\ref{eq:RamseyHahnProduct} and \ref{eq:RamseySingleTLF}) 
\begin{equation}
F(t) \approx \prod_{k=1}^n \exp{[-{\bar{v}}^2 t \gamma_k^{-1}/2]} = 
\exp{[-Z {\bar{v}}^2 t/2]},
\end{equation}
and the statistics of the curves $F(t)$ are determined by the statistics of the quantity
\begin{equation}
\quad Z=\sum_{k=1}^n \gamma_k^{-1}.
\end{equation}
It appears that for small TLF density $d\ll 1$, the probability distribution of $Z$ has the form $P_Z(z)\propto z^{-1+d}$, being controlled primarily by a single TLF with the largest value of $\gamma_k^{-1}$. This is shown via direct, although somewhat lengthy, calculation in Appendix~\ref{AppendixCohFidelity}, here we present only a simplified argument, explaining the essence of the phenomenon. 

The quantities $\gamma_k$, as well as $\gamma_k^{-1}$, all have identical log-uniform distributions (\ref{eq:PGamma}), such that their logarithms $\zeta_k=\ln{\gamma_k}$ are uniformly distributed between $\ln{\gamma_{m}}$ and $\ln{\gamma_{M}}$, in the region of the width 
$w=\ln{(\gamma_{M}/\gamma_{m})}$. When $n$ such quantities $\zeta_k$ are independently drawn from the region of width $w$, and ordered in an ascending order, the typical difference between two adjacent values of $\zeta_k$ is of the order $w/n=1/d$, which is large for $d\ll 1$. The corresponding quantities $\gamma_k^{-1}=\exp{(-\zeta_k)}$ are also ordered, in a descending manner, and each subsequent term is much smaller than its predecessor, being multiplied by a small factor of the order of $\exp{(-1/d)}$. As a result, the value of $Z$ is determined mostly by the single largest term in the sum $\sum_k\gamma_k^{-1}$, i.e.\ by the single smallest $\gamma_k$, which is distributed according to Eq.~\ref{eq:PGamma}. Therefore, $P_Z(z)\propto z^{-1}$ for $d\ll 1$, which is close (up to small correction $\sim d$) to the exact result $P_Z(z)\propto z^{-1+d}$
given in Appendix~\ref{AppendixCohFidelity}.

This simplified argument suggests that the validity of the Gaussian treatment is restored for dense baths with $d\gtrsim 1$: in that case, $n$ ordered quantities $\zeta_k$ are close to each other (the typical difference between two adjacent values is still of the order of $1/d$, which is now small), and the same is true for the quantities $\gamma_k^{-1}=\exp{(-\zeta_k)}$. In that case, the sum $Z$ consists of a large number $n$ of terms, and most of them are comparable to each other in magnitude, which is the situation where the central limit theorem is valid.

This argument also suggests that the total number $n$ of TLFs in a sparse bath is not a decisive parameter, as long as it is not small, and that only a few TLFs, which we will call exceptional TLFs from now on, with the smallest switching rates $\gamma$, determine the form of the Ramsey decay and the decay time $T_2^*$. Large sample-to-sample variability of the curves $F(t)$ comes from the fact that every bath parameter set $\mathcal{B}$ has its own exceptional TLFs, and their number is small, such that their specific parameters strongly fluctuate from one set $\mathcal{B}$ to another.
This conjecture is indeed supported by the numerical calculations presented below in Sec.~\ref{sec:excTLF}.

\begin{figure}[tbp!]
\centering
\includegraphics[width=1\linewidth]{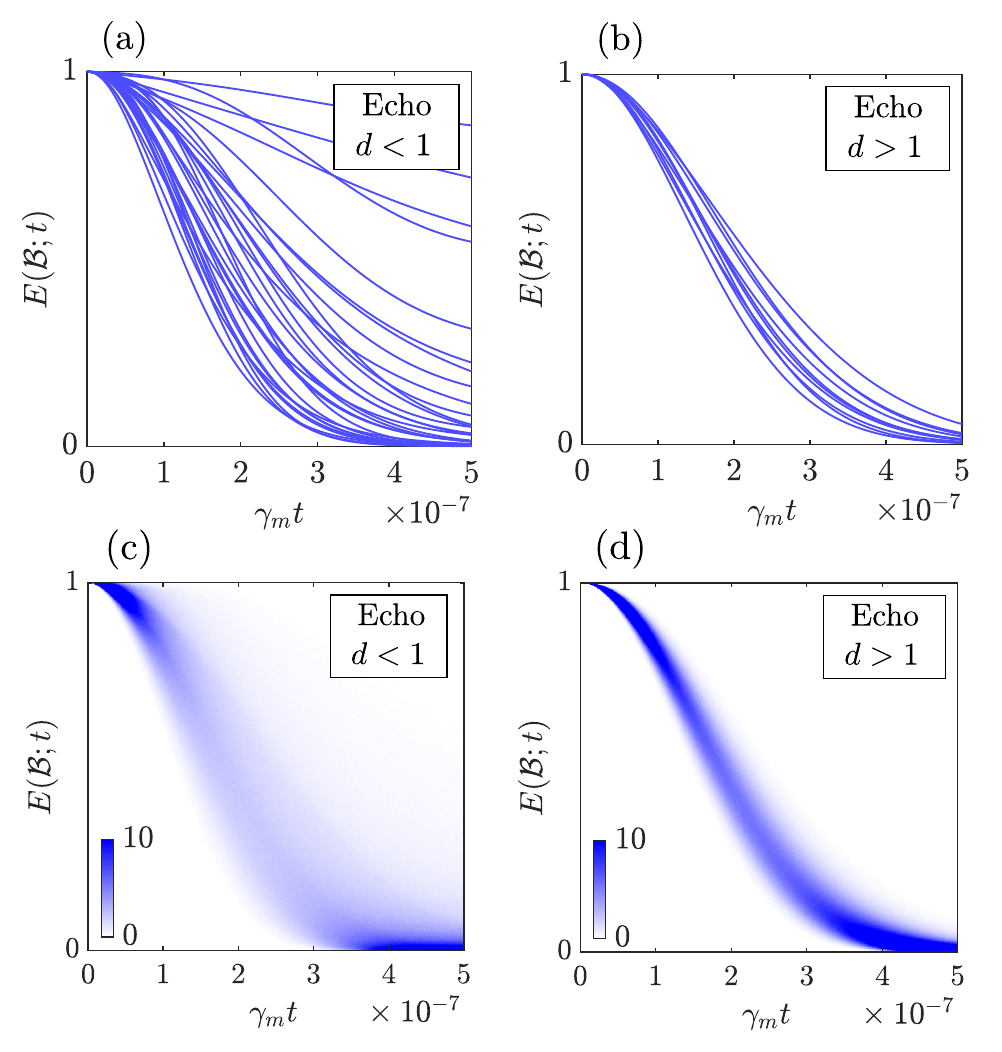}
\caption{Echo signal decay for sparse versus dense baths. The ratio $\gamma_M/\gamma_m=10^{12}$ is fixed in all panels. {\bf (a)} Samples of the echo decay curves $E(t)$ 
for sparse baths with $n=16$ TLFs, corresponding to $d\approx0.58$; the coupling strengths are the same for all TLFs, $v_k=\bar{v}=\gamma_m\cdot 10^7$. Thirty randomly chosen samples are shown. {\bf (b)} Samples of the echo decay $E(t)$ 
for dense baths with $n=100$ and $d\approx 3.62$; the coupling strengths are the same for all TLFs, $v_k=\bar{v}=4\gamma_m\cdot 10^6$, 
such that the variance $\beta(\mathcal{B})$ of the field $B(t)$ is the same as in panel (a). Ten randomly chosen samples are shown in order to make the individual curves clearly visible. 
{\bf (c)},{\bf (d)}: Two-dimensional histograms of the estimated p.d.f.'s $P_E(E,t)$ of the echo decay curves, presented in a qualitative form, for sparse baths, panel {\bf (c)}, and for dense baths, panel {\bf (d)}. The p.d.f.'s are obtained using $M=10^5$ bath parameter samples, and are calculated and presented in the same way as in Fig.~\ref{fig:Fig2}(c),(d).} 
\label{fig:Fig3}
\end{figure}

Meanwhile, it is interesting to explore similar qualitative difference between sparse and dense baths in the case of the echo decay. In the same manner as above, we assume that the qubit is prepared in the initial state with $m_x=1$, $m_y=m_z=0$, and interacts  with many TLFs as described by the Hamiltonian (\ref{eq:Ham}). After a period of free evolution of duration $\tau$, a refocusing $\pi$-pulse is applied to the qubit, and the system evolves for another time period of duration $\tau$; the value of $m_x(t=2\tau)$ is measured \cite{slichter} at $t=2\tau$. 
The shape of the echo decay for a qubit coupled to a single TLF is also known \cite{Paladino2002,Zhidomirov1969,deSousa2003,Galperin2006,RamonNonGauss2015},
\begin{equation}
\label{eq:HahnSingleTLF}
e(\gamma,v;t) = \mathrm{e}^{-\gamma t}\left(\frac{\gamma^2}{\alpha^2}\cosh{\alpha t} + \frac{\gamma}{\alpha} \sinh{\alpha t} - \frac{v^2}{\alpha^2}\right),
\end{equation}
with $\alpha=(\gamma^2 - v^2)^\frac{1}{2}$, and the echo decay for a bath of $n$ independent TLFs is a product of the individual decay factors 
\begin{equation}
m_x(t=2\tau)\equiv E(\mathcal{B};t)= \prod_{k=1}^n e_k (t),
\end{equation}
where $e_k(t)\equiv e(\gamma_k,v_k;t)$.

The echo decay factor for a slow TLF with $\gamma\ll v$ has the form $e(\gamma,v;t)\approx\exp{(-\gamma t)}$. For a fast TLF with $\gamma\gg v$, just like in the case of the Ramsey experiment, the decay mostly happens at 
$t\gtrsim \gamma^{-1}$, and the echo decay factor has the same form, $e(\gamma,v;t)\approx f(\gamma,v;t)\approx\exp{[-v^2 t \gamma^{-1}/2]}$.
Therefore, when the qubit is coupled to a bath of many TLFs with similar coupling strengths $v_k\sim \bar{v}$, the slow fluctuators with very small $\gamma$, as well as very fast TLFs with very large $\gamma$, contribute little to the overall echo signal decay, and the form of $E(\mathcal{B};t)$ is mostly controlled by the TLFs with $\gamma_k\sim\bar{v}$. 
Therefore, for investigating the echo decay in the physically relevant and interesting regime, we need to consider baths where $\bar{v}\sim\sqrt{\gamma_m\cdot\gamma_M}$, such that all three types of TLFs, with $\gamma_k\ll\bar{v}$, $\gamma_k\sim\bar{v}$ and $\gamma_k\gg\bar{v}$,  are present.

The results of such simulations are presented in Fig.~\ref{fig:Fig3}. Similarly to Fig.~\ref{fig:Fig2}, there is a stark contrast between sparse and dense baths. Variability of the echo decay curves $E(t)$ is very high for sparse baths with $d\ll 1$, and the individual curves differ greatly. For dense baths, the curves $E(t)$ are very close to each other, and the p.d.f.\ $P_E(E,t)$ is sharply concentrated around a single curve. This happens in spite of the fact that all baths, whether sparse or dense, have very close noise power spectra $S(\mathcal{B}_m;\omega)$, so that the conventional approach 
\cite{Cywinski2008prb,Schriefl2006,Szankowski2017,Galperin2006,Bergli2009,CywinskiWitzelDS2009,paladino2014rmp} is not applicable to the case of the echo decay caused by sparse baths.

These results suggest the possibility that the echo decay $E(t)$ is also controlled by only one or a few important TLFs, analogously to the situation with the Ramsey decay $F(t)$. However, in the case of echo such exceptional fluctuators can not be the slowest (with $\gamma_k\ll\bar{v}$) nor the fastest (with $\gamma_k\gg\bar{v}$), because TLFs of both these types contribute little to the echo decay. It is logical to assume that the shape of $E(t)$ is controlled by the TLFs which are the closest to satisfying the condition $\gamma_k\sim\bar{v}$. The results presented below in Sec.~\ref{sec:excTLF} confirm this conjecture. Therefore, large sample-to-sample variation of the curves $E(t)$ is also due to the fact that the number of such exceptional TLFs is small, and their parameters significantly fluctuate from one set  $\mathcal{B}$ to another.

Note that many previous works exploring non-Gaussian behavior incorrectly assume that a bath of many TLFs, coupled equally strongly to the qubit, automatically enters Gaussian regime and can be treated in a central limit theorem-like manner. Our arguments above show that this expectation is too naive, and more care is required.

\section{The role of individual fluctuators in dephasing of a qubit}
\label{sec:excTLF}

We have established in the previous section that the Ramsey decay, in the case of weakly coupled TLFs, is governed by only a few exceptional TLFs with the smallest values of the switching rate $\gamma$. We have also conjectured that the echo decay is also controlled by a few exceptional TLFs, whose switching rates $\gamma$ are close to the coupling $\bar{v}$.
A very straightforward way to check, whether these statements are correct, is to remove these exceptional TLFs, and analyze the changes in the decay curves $F(t)$ and $E(t)$.

\begin{figure}[tbp!]
\centering
\includegraphics[width=\linewidth]{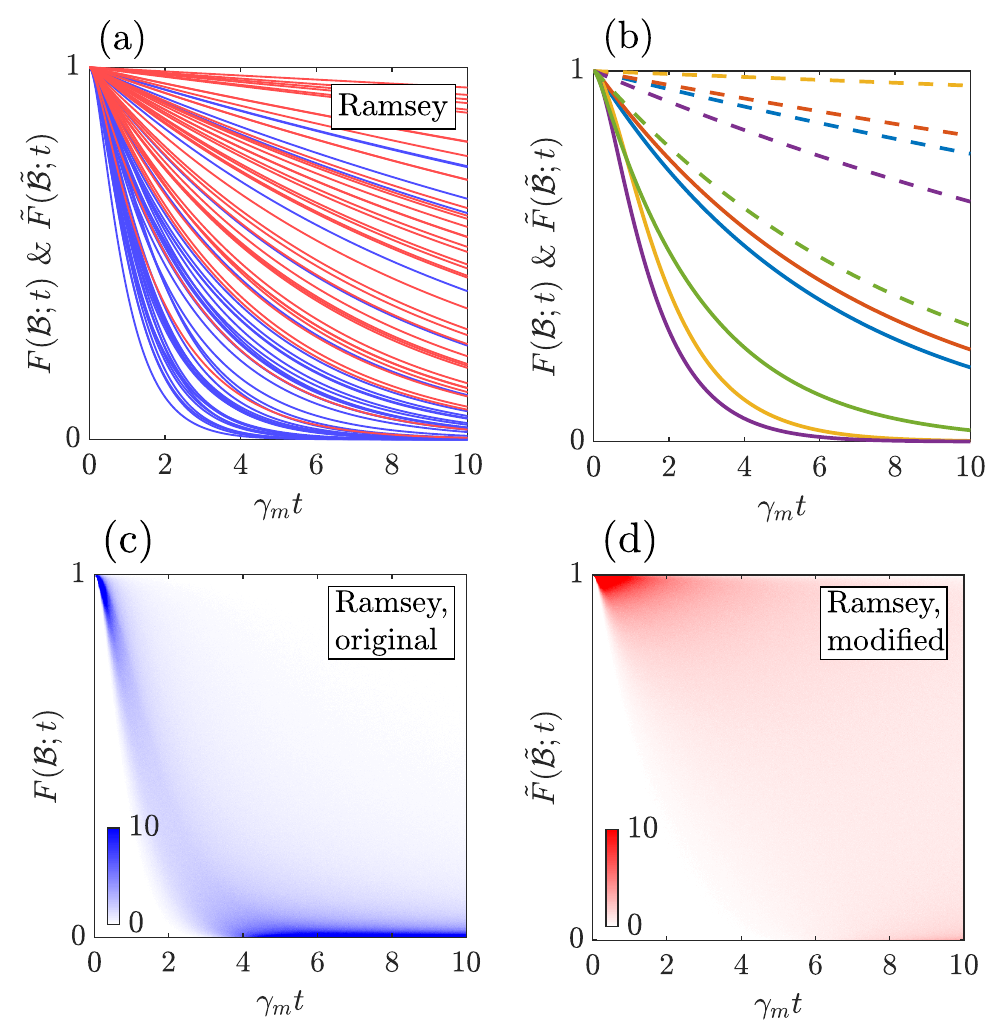}
\caption{Drastic changes in the Ramsey decay due to removal of two exceptional defects from sparse bath. All panels correspond to baths with the ratio $\gamma_M/\gamma_m=10^5$, containing $n=12$ TLFs, such that $d\approx 1.04$; the coupling strengths are the same for all TLFs, $v_k=\bar{v}=\gamma_m$.
{\bf (a)} Fourty randomly chosen samples $F_m(t)$, obtained with the original sets $\mathcal{B}_m$, are shown as blue lines; the corresponding curves $\tilde{F}_m(t)$, obtained with the modified parameter sets $\tilde{\mathcal{B}}_m$, are shown as red lines. Drastic slowdown of the Ramsey decay upon removal of two exceptional TLFs is clearly seen.
{\bf (b)} Comparison of the individual Ramsey curves obtained with original (solid lines) and modified (dashed lines) bath parameter sets. The curves   $F_1(t),\dots F_5(t)$ are color-coded, each color denoting one parameter set, ${\mathcal{B}}_1$ to ${\mathcal{B}}_5$. 
Each parameter set is modified by removal of two exceptional TLFs, producing the modified sets $\tilde{\mathcal{B}}_1$ to $\tilde{\mathcal{B}}_5$, and the corresponding curves $\tilde{F}_1(t),\dots \tilde{F}_5(t)$ are plotted with dashed lines, such that the curve $\tilde{F}_m(t)$ for every $m=1,\dots 5$ has the same color as the original ${F}_m(t)$. Comparison between the solid and the dashed lines of the same color shows drastic slowdown of the Ramsey decay for each individual bath sample. 
{\bf (c)}, {\bf (d)}: 
Two-dimensional histograms of the estimated p.d.f.\ $P_F(F,t)$ of the original Ramsey curves $F(t)$, panel {\bf (c)}, and of the estimated p.d.f.\  $\tilde{P}_F(\tilde{F},t)$
of the modified curves $\tilde{F}(t)$, panel {\bf (d)}. 
The estimated p.d.f.\ $P_F(F,t)$ was obtained from the set of Ramsey curves $F_m(t)$ ($m=1,\dots M$), which were calculated using the original parameter sets $\mathcal{B}_m$. Then every parameter set was modified by removing two exceptional TLFs, and the p.d.f.\ $\tilde{P}_F(\tilde{F},t)$ was obtained from  the Ramsey curves $\tilde{F}_m(t)$, which were calculated using the modified sets $\tilde{\mathcal{B}}_m$. Both histograms used $M=10^5$ samples of the bath parameter sets; both estimated p.d.f.'s are produced and presented in the same manner as in Fig.~\ref{fig:Fig2}(c),(d).}
\label{fig:Fig4}
\end{figure}

For that, we generate a large number $M$ of samples of the bath parameter sets $\mathcal{B}_m$, $m=1,\dots M$. Each set $\mathcal{B}_m$ is modified by removing two exceptional TLFs, producing the corresponding modified parameter set $\tilde{\mathcal{B}}_m$. The curve $F_m(t)\equiv F({\mathcal{B}}_m,t)$, calculated using the original set $\mathcal{B}_m$, and the curve $\tilde{F}_m(t)\equiv F(\tilde{\mathcal{B}}_m,t)$, calculated using the corresponding modified set $\tilde{\mathcal{B}}_m$, are compared with each other, both individually (separately for every $m$) and statistically (comparing the ensembles of $F_m(t)$ and $\tilde{F}_m(t)$). 

For dense baths, the decay curves $\tilde F_m(t)$ for every $m$ remain practically the same as the corresponding $F_m(t)$ curves, and the same happens for the echo decay curves $E_m(t)$ and $\tilde{E}_m(t)$. This is precisely what is expected in a Gaussian regime, where the effect of any individual TLF is small. However, for sparse baths, such a minor modification of the bath parameters leads to significant changes.

Fig.~\ref{fig:Fig4} shows substantial difference in the Ramsey decay curves $F(t)$, occurring after removal of two exceptional fluctuators. As seen in Fig.~\ref{fig:Fig4}(a), the curves $F_m(t)$ change substantially upon modification, and the modified curves $\tilde{F}_m(t)$ decay much slower than the original ones. Fig.~\ref{fig:Fig4}(b) demonstrates that this happens for every single parameter set: each of the color-coded original curves $F_m(t)$ undergoes such changes, and for every instance $m$, the function $\tilde{F}_m(t)$ decays much slower than the corresponding $F_m(t)$.

This trend is maintained statistically, for the ensemble of $M=10^5$ original baths ${\mathcal{B}}_m$ ($m=1,\dots M$) and the corresponding modified baths $\tilde{\mathcal{B}}_m$. This is shown in Fig.~\ref{fig:Fig4}(c) and (d), which compare the estimated p.d.f.\ $P_F(F,t)$ of the original Ramsey curves $F(t)$ with the estimated p.d.f.\  $\tilde{P}_F(\tilde{F},t)$ of the modified curves $\tilde{F}(t)$. The estimated p.d.f.'s were obtained as described above in Fig.~\ref{fig:Fig2}(c),(d) and in the corresponding text. Upon removal of exceptional TLFs, the density of the Ramsey decay curves $F(t)$ moves upwards very noticeably, demonstrating significant slowdown of the Ramsey decays.

\begin{figure}[tbp!]
\centering
\includegraphics[width=\linewidth]{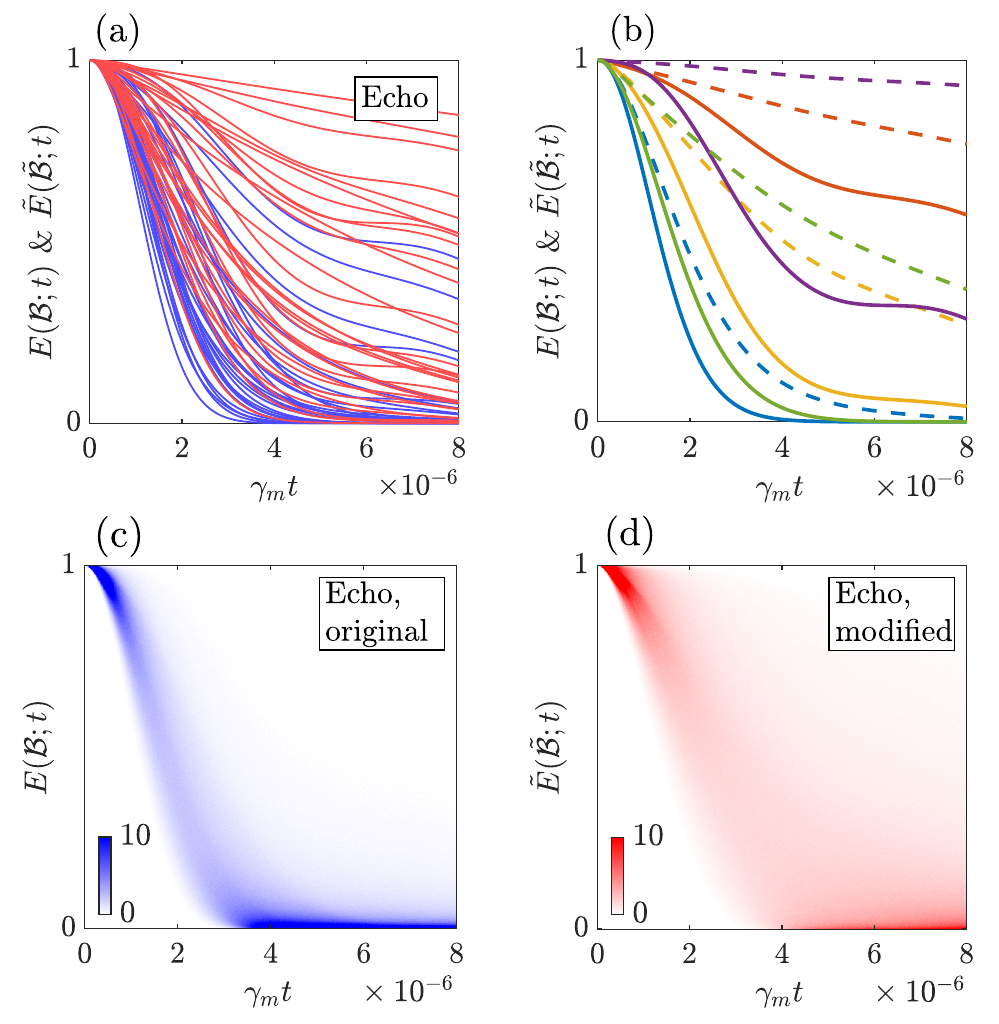}
\caption{Drastic changes in the echo decay upon removal of two exceptional defects from sparse bath. All panels correspond to baths with the ratio $\gamma_M/\gamma_m=10^{12}$, containing $n=20$ TLFs, such that $d\approx0.72$; the coupling strengths are the same for all TLFs, $v_k=\bar{v}=\gamma_m$. The bath parameter sets were modified by removing two exceptional TLFs, whose switching rate $\gamma_k$ is the closest to the value $\bar{v}$.
{\bf (a)} Thirty randomly chosen samples $E_m(t)$, obtained with the original sets $\mathcal{B}_m$, are shown as blue lines; the corresponding curves $\tilde{E}_m(t)$, obtained with the modified parameter sets $\tilde{\mathcal{B}}_m$, are shown as red lines. Drastic slowdown of the echo decay upon removal of two exceptional TLFs is clearly visible.
{\bf (b)} Comparison of the individual echo decay curves obtained with original (solid lines) and modified (dashed lines) bath parameter sets. The curves   $E_1(t),\dots E_5(t)$, obtained using the parameter sets ${\mathcal{B}}_1\dots{\mathcal{B}}_5$, are color-coded. The corresponding curves $\tilde{E}_1(t),\dots \tilde{E}_5(t)$, obtained from the modified sets $\tilde{\mathcal{B}}_1\dots\tilde{\mathcal{B}}_5$. 
are plotted with dashed lines of the corresponding color, as in Fig.~\ref{fig:Fig4}(b). Comparison between the solid and the dashed lines of the same color shows drastic slowdown of the echo decay for each individual bath sample. 
{\bf (c)}, {\bf (d)}: 
Two-dimensional histograms of the estimated p.d.f.\ $P_E(E,t)$ of the original echo curves $E(t)$, panel {\bf (c)}, and of the estimated p.d.f.\  $\tilde{P}_E(\tilde{E},t)$
of the modified curves $\tilde{E}(t)$, panel {\bf (d)}. The p.d.f.'s were obtained in the same way as in Fig.~\ref{fig:Fig4}(c),(d).
Both histograms used $M=10^5$ samples of the bath parameter sets.}
\label{fig:Fig6}
\end{figure}

The same conclusions hold for the echo decays, as seen in Fig.~\ref{fig:Fig6}, where we show the changes in the echo decay curves $E(t)$ caused by removal of two exceptional TLFs from the bath. In the same manner as it was done above for the Ramsey decay curves, we generate a large number $M$ of samples of the bath parameter sets $\mathcal{B}_m$, $m=1,\dots M$; we modify each $\mathcal{B}_m$, producing the corresponding modified set $\tilde{\mathcal{B}}_m$. The curves $E_m(t)$, calculated using the original set $\mathcal{B}_m$, and the curves $\tilde{F}_m(t)$, calculated using the corresponding modified set $\tilde{\mathcal{B}}_m$, are compared with each other, both individually (separately for several randomly chosen values of $m$) and statistically (comparing the ensembles of $E_m(t)$ and $\tilde{E}_m(t)$). The comparison shows that the echo decay for each individual bath greatly slows down upon removal of two exceptional TLFs, see Fig.~\ref{fig:Fig6}(a) and (b). This effect remains significant also at the level of large ensembles: as shown in Fig.~\ref{fig:Fig6}(c) and (d), substantial fraction of the p.d.f.'s $P_E(E,t)$ moves upwards, demonstrating substantial slowdown of the echo decay at the level of ensembles.
Note that the decisive role of similar individual TLFs has also been recognized in earlier studies \cite{Galperin2006,Bergli2009,Paladino2002}, although for other bath models and regimes. Agreement with these previous results provides assurance that our results remain valid in a sufficiently broad range of physical situations.

\section{Improvements in the qubit properties upon removal of exceptional TLFs}
\label{sec:improve}

\begin{figure}[ht!]
\centering
\includegraphics[width=1\linewidth]{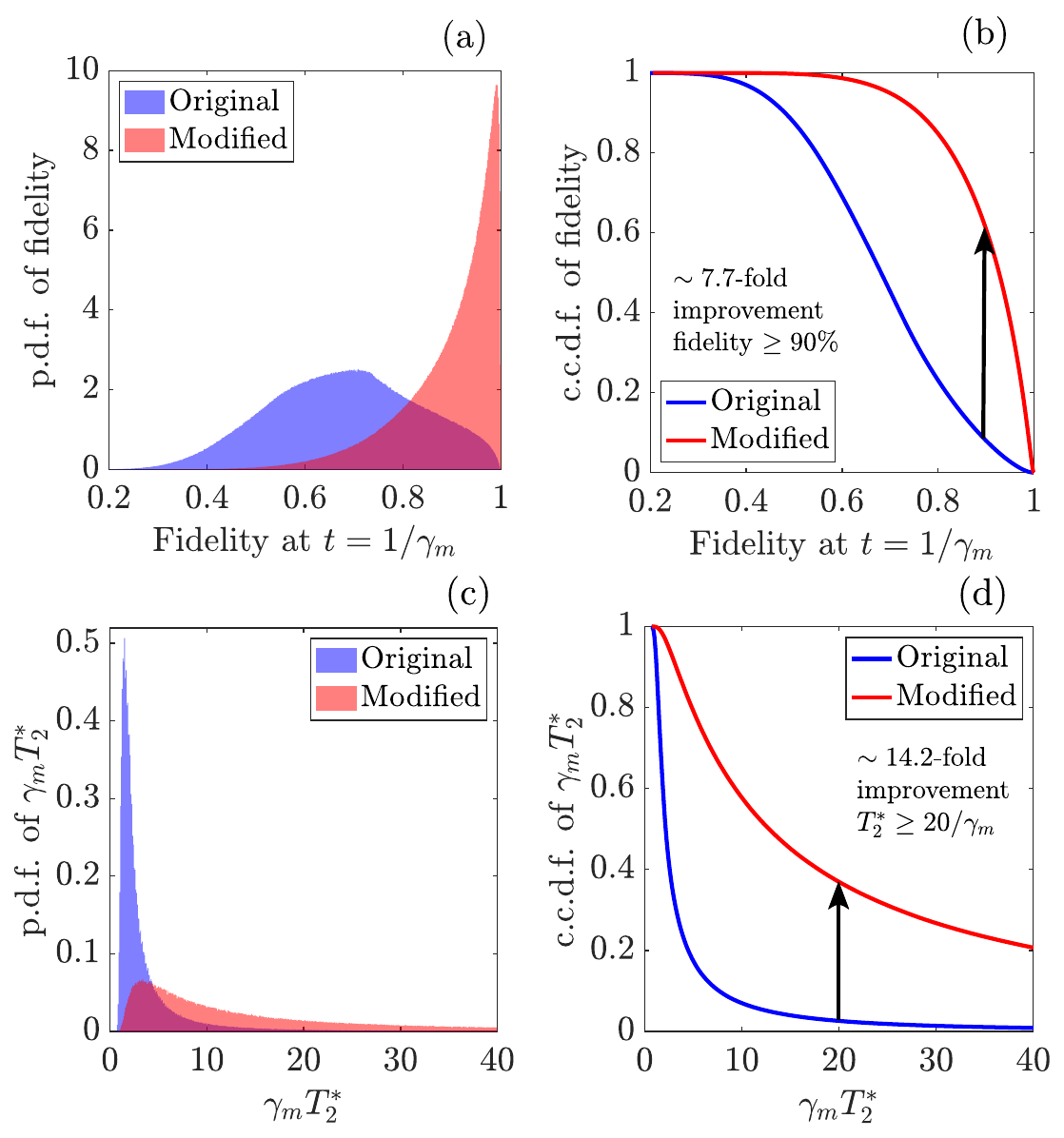}
\caption{Improvements in the qubit quality, quantified by two metrics, the Ramsey decay time $T_2^*$ and the qubit fidelity $\mathcal{F}=F(t_0)$, where $t_0=1/\gamma_m$. The same parameters as in Fig.~\ref{fig:Fig4} were used to produce the plots. {\bf (a)} Estimated p.d.f.'s of the fidelity $\mathcal{F}$ for the original (blue) and modified (red) baths. The p.d.f.\ moves to the region of larger fidelities upon removal of the exceptional TLFs, showing improvements in the fidelity. $M=10^7$ samples of the bath parameters were used; the bin width $\delta=1.75\cdot 10^{-3}$ for original baths and $\delta=1.606\cdot 10^{-3}$ for modified baths. 
{\bf (b)} Estimated c.c.d.f.'s of the fidelity $\mathcal{F}$ for the original (blue line) and modified (red line) baths. The red line is noticeably shifted upwards and to the right, to the region of higher fidelities, in comparison with the blue line. The fraction of the high-fidelity qubits, with 
$\mathcal{F}>0.9$, increases by a factor of $7.7$ upon removal of two exceptional TLFs. Bin parameters are the same as in panel (a).
{\bf (c)} Estimated p.d.f.'s of the decay time $T_2^*$ for the original (blue) and modified (red) baths. $M=10^6$ samples were used; the bin width is $8.28\cdot 10^{-2}/\gamma_m$ for original baths and $8.46\cdot 10^{-2}/\gamma_m$ for modified baths. 
{\bf (d)} Estimated c.c.d.f.'s of the decay time $T_2^*$ for the original (blue line) and modified (red line) baths. Bin parameters are the same as in panel (c).
Statistics of the time $T_2^*$ shows the same trends as the fidelity $\mathcal{F}$. In particular, the fraction of the high-quality qubits, with the decay time $T_2^*\ge 20/\gamma_m$, increases by a factor of $14.2$ upon removal of two exceptional TLFs.} 
\label{fig:Fig5} 
\end{figure}

The results of the previous Section suggest that it might be possible to drastically improve the coherence properties of a qubit by eliminating only a few most offending TLFs, without radical changes in the device or in the fabrication process. For instance, it has been demonstrated \cite{Meyer2023electrical,Lisenfeld2023enhancing,RyuKangDevitalizeDefects22} that individual defects can be deactivated in the already fabricated devices by adjusting the working point parameters. For the devices where the defects make up a sparse bath, similar to the one studied here, such deactivation may be effective. However, we do not discuss here, how realistic is this possibility: this decision is to be made by experimentalists, material scientists and engineers. We focus on theoretical investigation of great potential benefits of this option.

\begin{figure}[t!]
\centering
\includegraphics[width=\linewidth]{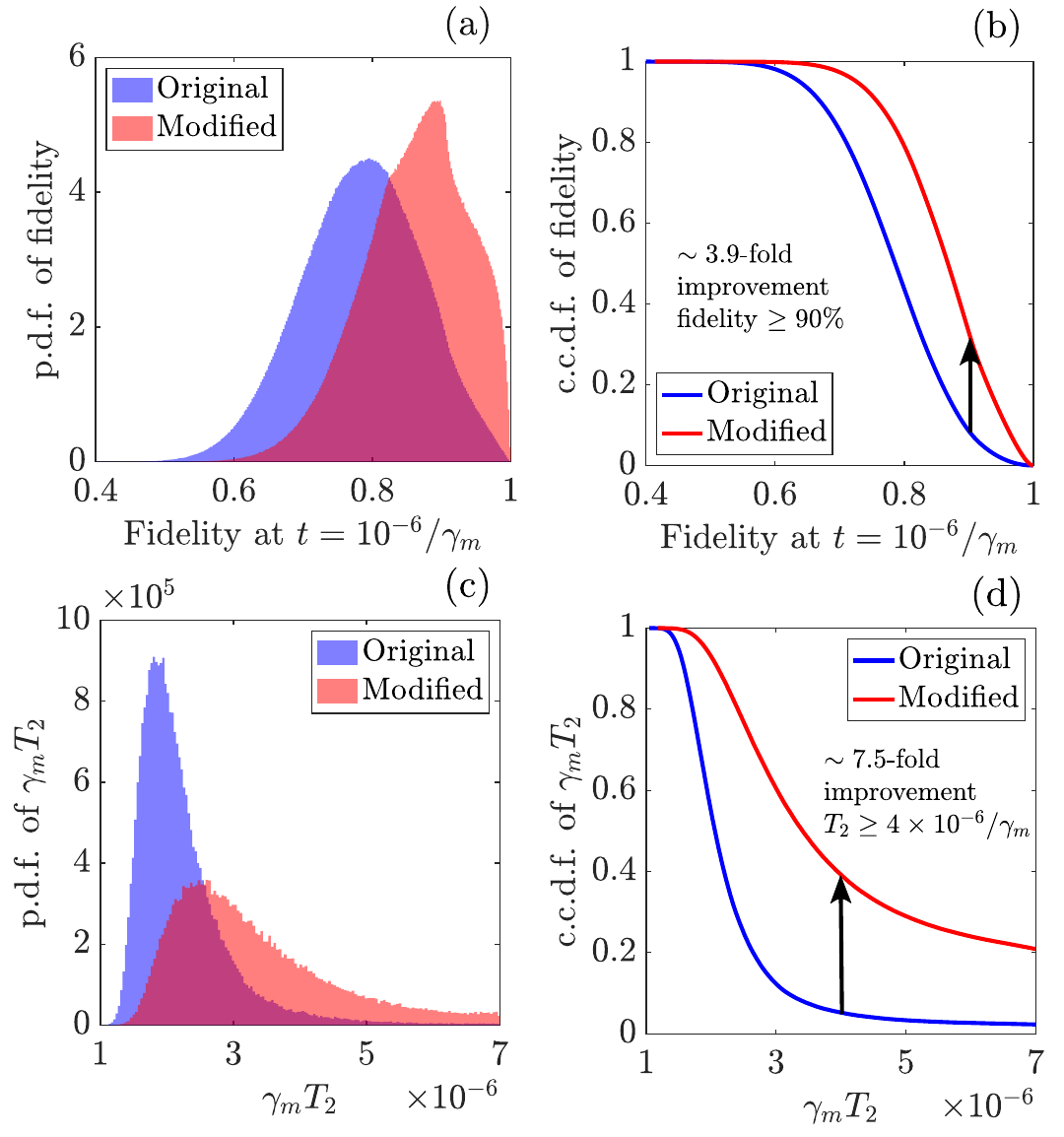}
\caption{Improvements in the qubit quality, quantified by two metrics, the echo decay time $T_2$ and the echo fidelity $\mathcal{E}=E(t_0)$, where $t_0=10^{-6}/\gamma_m$. The same parameters as in Fig.~\ref{fig:Fig6} were used to produce the plots. {\bf (a)} Estimated p.d.f.'s of the fidelity $\mathcal{E}$ for the original (blue) and modified (red) baths. The p.d.f.\ moves to the region of larger fidelities upon removal of the exceptional TLFs, showing improvements in the fidelity. 
$M=10^7$ samples of the bath parameters were used, with the data distributed over 300 bins of the width $\delta=2.112\cdot 10^{-3}$ for original baths and $\delta=2.02\cdot 10^{-3}$ for modified baths.
{\bf (b)} Estimated c.c.d.f.'s of the fidelity $\mathcal{E}$ for the original (blue line) and modified (red line) baths. The red line noticeably shifts to the region of higher fidelities, in comparison with the blue line. The fraction of the high-fidelity qubits, with $\mathcal{E}>0.9$, increases by a factor of $3.9$ upon removal of two exceptional TLFs. Bin parameters are the same as in panel (a).
{\bf (c)} Estimated p.d.f.'s of the echo decay time $T_2$ for the original (blue) and modified (red) baths. $M=2\cdot 10^5$ samples were used, with the data distributed over 200 bins of the width $\delta=3.49\cdot 10^{-8}/\gamma_m$ for original baths and $\delta=3.92\cdot 10^{-8}/\gamma_m$ for modified baths.
{\bf (d)} Estimated c.c.d.f.'s of the echo decay time $T_2$ for the original (blue line) and modified (red line) baths. Bin parameters are the same as in panel (c). Statistics of $T_2$ shows the same trends as the fidelity $\mathcal{F}$. The fraction of high-quality qubits, with the decay time $T_2\ge 4\times10^{-6}/\gamma_m$, increases by a factor of $7.5$ upon removal of two exceptional TLFs.} 
\label{fig:Fig7}
\end{figure}

We quantify the improvements by analyzing two metrics. The first one is the Ramsey decay time $T_2^*$, defined as the time when the Ramsey decay curve reaches the value 1/e. The second one is the Ramsey decay fidelity $\mathcal{F}$, defined as the value of the Ramsey decay function at a certain small time $t_0$, i.e.\ $\mathcal{F}\equiv F(t_0)$. In order to quantify the improvements caused by eliminating the exceptional defects, we calculate the p.d.f.'s and the complementary cumulative distribution functions (c.c.d.f.'s) for both quantities, $T_2^*$ and $\mathcal{F}$; the c.c.d.f.\ $\mathcal{C}_X$ of a quantity $X$, which is distributed with the p.d.f.\ $P_X(x)$, is defined as $\mathcal{C}_X(x) = \int_x^\infty P_X(u)du$. 

Fig.~\ref{fig:Fig5} shows, how p.d.f.'s and c.c.d.f.'s of both metrics, $T_2^*$ and $\mathcal{F}$, change upon modification of the bath parameter sets.  
The plots were produced by sampling $M=10^7$ original sets $\mathcal{B}_m$, and removing two exceptional TLFs from each set, thus obtaining the modified set $\tilde{\mathcal{B}}_m$ for every $m=1,\dots M$. The Ramsey decay curves $F_m(t)$ and $\tilde{F}_m(t)$ were calculated, and the values of $T_2^*$ and $\mathcal{F}$ were extracted for each $m$. These values were used for estimating p.d.f.'s and c.c.d.f.'s of the two metrics, by distributing the results into 500 bins of the width $\delta$; the number of counts in each bin was divided by the factor $M\cdot\delta$ in order to normalize the estimated p.d.f.'s $P_X(u)$ as $\int P_X(u) du = 1$ for both $X=T_2^*$ and $X=\mathcal{F}$.

As seen from Fig.~\ref{fig:Fig5}(a) and (c), both metrics are greatly improved after removing two exceptional TLFs. Specifically, the fraction of high-fidelity qubits, with $\mathcal{F}>0.9$, increases by a factor of 7.7, as shown in Fig.~\ref{fig:Fig5}(b), and the fraction of qubits with $T_2^*\ge 20/\gamma_m$ improves by a factor of 14.

The same conclusions hold if we quantify the echo decay in a similar way, using the echo decay time $T_2$ and the echo fidelity $\mathcal{E}$ as metrics; the echo fidelity is defined as the value of the echo decay function at certain small time $t_0$, i.e.\ $\mathcal{E}\equiv E(t_0)$.
Fig.~\ref{fig:Fig7} shows, how p.d.f.'s and c.c.d.f.'s of both metrics change  after modification of the bath parameter sets. The data were binned, and the p.d.f.'s normalized, in the same manner as in Fig.~\ref{fig:Fig5}. Significant improvements, similar to those observed for the Ramsey decay, are clearly seen. 
Removal of two exceptional defects enhances the fraction of high-fidelity qubits, whose echo fidelity exceeds $0.9$ at time $t=10^{-6}/\gamma_m$, by a factor of $3.9$. The fraction of high-quality qubits, with the echo decay time  $T_2\ge4\times10^{-6}/\gamma_m$, increases by a factor of $7.5$.

\section{Sparse baths with finite dispersion of the coupling strengths}
\label{sec:dispersv}

Here we address the question of possible finite dispersion in the coupling strengths of the TLFs comprising the bath. An infinitely narrow distribution of coupling strengths, when all TLFs have the same values of the parameters $v_k$, is a crude approximation. It would be interesting to see, whether our results remain correct for a somewhat more realistic distribution of couplings. For this purpose, we consider normal distribution of the coupling strengths, characterized by the mean value $\bar{v}$ and the standard deviation $\sigma_v$.

\begin{figure}[tbp!]
\centering
\includegraphics[width=1\linewidth]{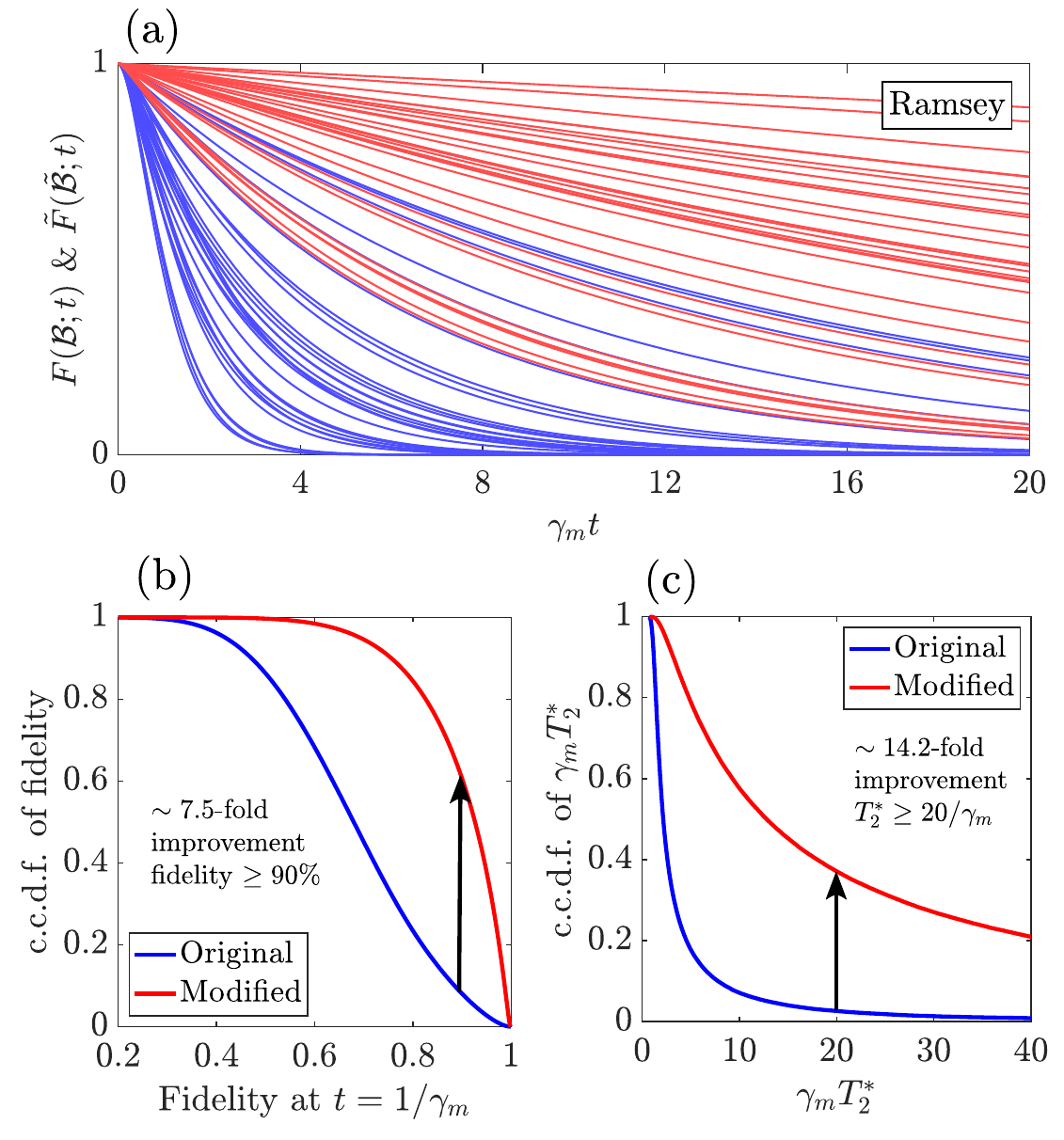}
\caption{Improvements in the Ramsey decay upon removal of two exceptional defects in the case of a normally distributed coupling strengths. The bath contains $n=12$ TLFs, and the ratio of the cutoff rates $\gamma_M/\gamma_m=10^5$, such that the TLF density $d\approx1.04$. The coupling strengths are normally distributed with the mean $\bar{v}=\gamma_m$ and the standard deviation $\sigma_v=\gamma_m/10$. 
{\bf (a)} Thirty samples of the Ramsey decay curves obtained using the original (blue curves) and the modified (red curves) baths. Each bath sample was modified by removing two TLFs with the switching rates closest to $\bar{v}$, i.e.\ with the smallest rates $\gamma_k$. 
{\bf (b)} Estimated c.c.d.f.\ of the Ramsey fidelity $\mathcal{F}=F(t_0)$, where $t_0=1/\gamma_m$ for the original (blue line) and the modified (red line) baths. The fraction of high-fidelity qubits, with $\mathcal{F}>0.9$, increases by a factor of $7.5$ upon modification of the bath. 
{\bf (c)} Estimated c.c.d.f.\ of the Ramsey decay time $T_2^*$ for the original (blue line) and the modified (red line) baths. The fraction of high-quality qubits,  with $T_2^*\ge 20/\gamma_m$, increases by a factor of $14.2$ upon removal of the exceptional TLFs.} 
\label{fig:Fig8}
\end{figure}

However, if the sparse bath in question contains a TLF whose strength is too large, then the noise power spectrum $S(\mathcal{B},\omega)$ acquires a conspicuous Lorentzian peak at the corresponding frequency. Such a peak (or other equivalent signatures of an unusually strongly coupled TLF) would be visible in experiments, and the noise spectrum would noticeably deviate from the  $1/f$ form, assumed in this work. Such situations are realistic, but are outside of the scope of the present work, where we focus on the baths with reasonably smooth $1/f$ noise spectrum.

\begin{figure}[tbp!]
\centering
\includegraphics[width=\linewidth]{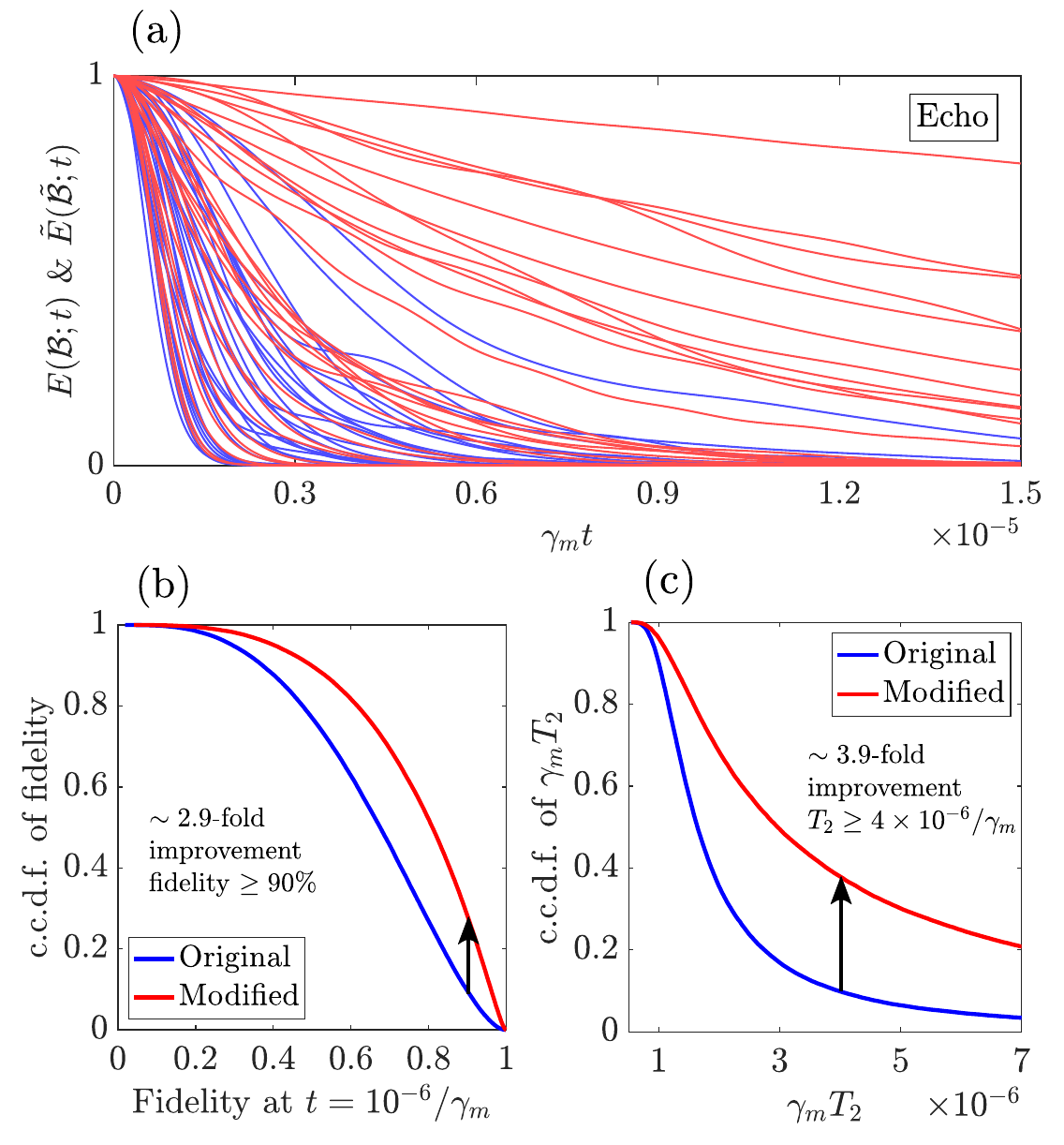}
\caption{Improvements in the echo decay upon removal of two exceptional defects in the case of a normally distributed coupling strengths. The bath contains $n=20$ TLFs, and the ratio of the cutoff rates $\gamma_M/\gamma_m=10^{12}$, corresponding to $d\approx 0.72$. The coupling strengths are normally distributed with the mean $\bar{v}=10^6\gamma_m$ and the standard deviation $\sigma_v=\bar{v}$. Each bath parameter sample was modified by removing two TLFs with rates closest to $\bar{v}$. 
{\bf (a)} Thirty samples of the echo decay curves obtained using the original (blue curves) and the modified (red curves) baths. The slowdown of the echo decays is clearly seen.
{\bf (b)} Estimated c.c.d.f.\ of the echo fidelity $\mathcal{E}=E(t_0)$, where $t_0=110^{-6}/\gamma_m$, for the original (blue line) and the modified (red line) baths. The fraction of high-fidelity qubits, with $\mathcal{F}>0.9$, increases by a factor of $2.9$ upon modification of the bath. 
{\bf (c)} Estimated c.c.d.f.\ of the echo decay times $T_2$ for the original (blue line) and the modified (red line) baths. The fraction of high-quality qubits, with $T_2\ge 4\times10^{-6}/\gamma_m$, increases by a factor of $3.9$ upon removal of the exceptional TLFs.}
\label{fig:Fig9}
\end{figure}

In order to avoid such features in the noise spectrum, we impose the condition that the ratio of $\sigma_v$ to $\bar{v}$ is much smaller than the ratio of the cutoff rates $\gamma_M$ and $\gamma_m$, i.e.\
\begin{equation}
\label{eq:dispers}
\frac{\sigma_v}{\bar{v}} \ll \frac{\gamma_M}{\gamma_m}.
\end{equation}
Note that this condition does not preclude the distribution of the coupling strengths from being rather broad: depending on the specific values of  $\gamma_M$ and $\gamma_m$, this regime may include the situation of large dispersion in the coupling strengths, with ${\sigma_v}\sim {\bar{v}}$, such as the one addressed in Fig.~\ref{fig:Fig9}.

In this regime, we see the same qualitative features as above. Fig.~\ref{fig:Fig8}(a) shows high variability of the Ramsey decay curves, and the drastic slowdown of the Ramsey decay upon removal of two exceptional TLFs, whose switching rates are closest to the mean coupling value $\bar{v}$. Fig.~\ref{fig:Fig8}(b) shows significant increase in the fraction of the qubits with high fidelity ($\mathcal{F}>0.9$), by a factor of 7.5 in comparison with the original baths. Fig.~\ref{fig:Fig8}(c) shows significant increase in the fraction of the high-quality qubits, with $T_2^*\ge 20/\gamma_m$, by factor of 14.2, upon removal of the exceptional TLFs. 
Similar features are seen in Fig.~\ref{fig:Fig9} for the echo decay and the corresponding metrics, the echo decay time $T_2$ and the echo fidelity $\mathcal{E}$.

\section{Dynamical decoupling of the qubit from sparse baths}
\label{sec:dd}

A very efficient way to protect the qubit from dephasing is to employ the dynamical decoupling (DD) technique, by applying a train of refocusing $\pi$ pulses to the qubit \cite{slichter,haeberlen,ViolaLloydDD,Cywinski2008prb,CywinskiWitzelDS2009,RamonNonGauss2015,paladino2014rmp}. A natural question arises about the dynamics of a qubit coupled to a sparse bath in the DD regime.

The calculation of the dephasing signal in that case can be performed in the same manner as for the Ramsey and the echo decays above. Assume that the qubit is prepared in the initial state with $m_x=1$, $m_y=m_z=0$, and undergoes dephasing under the action of the Hamiltonian (\ref{eq:Ham}). To be specific, let us focus on the Carr-Purcell (CP) decoupling protocol, with ideal instantaneous decoupling pulses being applied to the qubit at the moments of time $t_j=(2j-1)\tau$, with $j=1,2,\dots N$, such that a total of $N$ pulses is applied with the inter-pulse delay $2\tau$. The value $m_x(t)$ is measured at the moment of time $t= 2\,N\,\tau$, i.e.\ with the delay $\tau$ after the last refocusing pulse.

The shape of the CP signal decay for a qubit coupled to a single TLF is also known \cite{RamonNonGauss2015}, and has a form 
\begin{eqnarray}
\nonumber
c(\gamma,v;t) &=& \frac{1}{2}\,\mathrm{e}^{-\gamma t} 
\biggl[ 
\frac{\gamma^2\,\cosh{2\alpha\tau} - v^2}{\alpha\,\sqrt{\gamma^2 
\cosh^2{2\alpha\tau} - v^2}}\,\left(\mu_+^{N} - \mu_-^N\right) \\
  &+& \mu_+^{N} + \mu_-^N \biggr],
\label{eq:DDSingleTLF}
\end{eqnarray}
where
\begin{equation}
\mu_{\pm} = \frac{\gamma}{\alpha}\,\left[\sinh{2\alpha\tau} \pm \sqrt{\cosh^2{2\alpha\tau}-v^2/\gamma^2} \right].
\end{equation}
with $\alpha=(\gamma^2 - v^2)^\frac{1}{2}$. The CP signal decay curve for a bath of $n$ independent TLFs is a product of the individual decay factors 
\begin{equation}
C(\mathcal{B};t)= \prod_{k=1}^n c_k (t),
\end{equation}
where $c_k(t)\equiv c(\gamma_k,v_k;t)$.

\begin{figure}[tbp!]
\includegraphics[width=\linewidth]{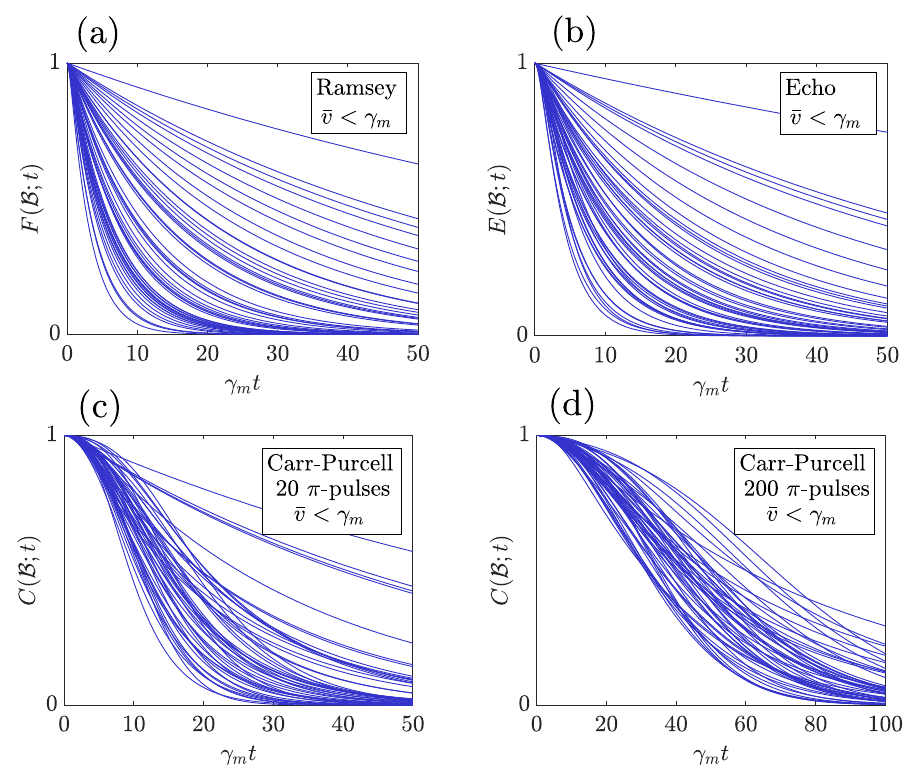}
\caption{Qubit decoherence for sparse baths in the weak coupling regime under different qubit control sequences. {\bf (a)} Ramsey decay, {\bf (b)} Echo decay, {\bf (c)} CP decoupling sequence with 20 pulses, {\bf (d)} CP decoupling sequence with 200 pulses. Each panel shows the decay curves for the same thirty samples of the bath parameter sets $\mathcal{B}_m$, $m=1,\dots 30$, the samples were obtained with the same procedure as in all other figures. Each bath contains $n=10$ TLFs, the ratio $\gamma_M/\gamma_m=10^5$, corresponding to $d=0.87$. The coupling strengths are normally distributed with the mean $\bar{v}=0.5\cdot\gamma_m$ and the standard deviation $\sigma_v=0.01\cdot\gamma_m$.
The decay curves show that significant increase in the number of pulses leads to only moderate improvement in the decoherence rate, but significantly reduces the sample-to-sample variability.}
\label{fig:Fig13} 
\end{figure}

The corresponding decoherence curves for sparse baths in the weak coupling regime, with $\bar{v}<\gamma_m$ and small dispersion $\sigma_v$, are shown in Fig.~\ref{fig:Fig13}. The results are very surprising: improvements in the decoherence time are rather modest, even for 200 decoupling pulses. Instead, the sample-to-sample variability, so pronounced in the Ramsey and in the echo decays, is greatly reduced by the CP sequence. Besides, the overall decay curve acquires Gaussian form, in contrast with the Ramsey and the echo decays, having a well-pronounced exponential form.

To understand this strange behavior, we notice that in the limit of very large number of pulses $N$, shown in Fig.~\ref{fig:Fig13}(d), and small couplings $v_k\ll \gamma_k$, the role of the terms $\mu_-^N$ is negligible, while the factor multiplying the term $\mu_+^N$ inside the square brackets of Eq.~\ref{eq:DDSingleTLF} is close to 2. Thus, the individual decoherence factor for a single TLF has an approximate form
\begin{eqnarray}
c(\gamma,v;t) &\approx & \mathrm{e}^{-\gamma t} \mu^N,\\ \nonumber
\ln{\mu} &\approx & 2\gamma\tau - v^2\,\frac{2\gamma\tau - 
\mathrm{tanh}(2\gamma\tau)}{2\,\gamma^2},
\end{eqnarray}
such that 
\begin{eqnarray}
c(\gamma,v;t) & \approx & \exp{\left[-\frac{v^2 t^2}{2 N^2}\,Q(\gamma\tau)\right]},\\ \nonumber
Q(\gamma\tau) &=& \frac{2\gamma\tau - \mathrm{tanh}(2\gamma\tau)}{\gamma^2 \tau^2}.
\end{eqnarray}
By taking the product over all TLFs, we obtain the result
\begin{equation}
\label{eq:sCP}
C(\mathcal{B};t)= \exp{\left[-\frac{{\bar{v}}^2 t^2}{2 N^2}\,\sum_k Q(\gamma_k\tau)\right]}
\end{equation}
where we took into account that the dispersion of the coupling strengths is small, $\sigma_v\ll\bar{v}$, such that $v_k\approx\bar{v}$.

Now let us look closer at the sum $\sum_k Q(\gamma_k\tau)$ in the situation of a sparse bath, where $\gamma_k$ are distributed log-uniformly over a wide range of values, such that $\gamma\tau$ varies in a broad range. We follow the same approach as in Sec.~\ref{sec:three}, introducing the quantities $\zeta_k=\ln{\gamma_k}$, which are distributed uniformly over the range $[\ln{\gamma_m},\ln{\gamma_M}]$. In terms of the parameters $\zeta$, the function in question is expressed as $Q(\gamma\tau) = Q_0(\zeta+\theta)$, where we introduced the quantity $\theta=\ln{\tau}$ and the function
\begin{equation}
Q_0(x) = \frac{2\,\mathrm{e}^{x} - \mathrm{tanh}(2\mathrm{e}^{x})}{\mathrm{e}^{2x}}.
\end{equation}
The function $Q_0(x)$ has a bell-like shape, with a broad peak stretching from  $\sim -2$ to $\sim 2$, while the quantity $\zeta$ is distributed uniformly over the range $w=\ln{10^5}\approx 12$ (the value used in Fig.~\ref{fig:Fig13}). That is, more than 1/3 of all TLFs contribute to decoherence with comparable effect. At the same time, the value of $\theta$ changes weakly (logarithmically) as $t$ increases, and does not significantly shift the relevant range of $\zeta$.
As a result, we restore, to a noticeable extent, the Gaussian regime of decoherence, when many TLFs contribute to decoherence with similar effect. 
Eq.~\ref{eq:sCP} demonstrates how, in this regime, the overall form of the signal decay acquires Gaussian form.
We note that similar effect was observed for a single TLF \cite{Cywinski2008prb,RamonNonGauss2015}, but the non-Gaussian features of a sparse bath of TLFs were not studied there. 

Detailed investigation of various regimes of the signal decay for a dynamically decoupled qubit interacting with a sparse bath of TLFs is beyond the scope of the present paper, and will be studied in a separate publication. However, it is important to spell out an essential observation made here, namely, that 
the behavior of the qubit coupled to a sparse bath of TLFs and subjected to a train of closely spaced control pulses is rather different from Ramsey or echo decay. In the latter cases, there are only one or two exceptional TLFs, which almost fully control the decoherence dynamics; these TLFs are determined by the relation between the switching rate $\gamma$ and the coupling strength $v$. In the case of dynamical decoupling, the effect of a given TLF, weakly coupled to the qubit, is determined instead by the relation between $\gamma$ and the delay $\tau$, and this effect, quantified by the function $Q_0(x)$, is not very selective, restoring Gaussian decoherence regime.

\section{Summary and conclusions}
\label{sec:sum}

In this work we investigated dynamics of dephasing of a qubit coupled to a bath of TLFs producing $1/f$ noise. We focused on the case of sparse bath, where the number $n$ of TLFs, although large, is still much smaller than the log-width $w=\ln{[\gamma_M/\gamma_m]}$ of the range $[\gamma_m,\gamma_M]$
of the switching rates $\gamma$, such that the density $d=n/w$ is small. We have shown that this bath possesses a number of very unconventional features. In order to separate the effect of the TLF density, we considered the baths where the coupling strengths of individual TLFs to the qubit are comparable, or have moderate dispersion (Eq.~\ref{eq:dispers} and Fig.~\ref{fig:Fig9}), such that the main (or sometimes the only) difference between different baths is the set of specific values of the switching rates of individual TLFs.

(i) We have demonstrated, both analytically and numerically, that the Gaussian regime of decoherence requires not only a large number $n\gg 1$ of TLFs in the bath, as one might expect, but also large density $d$. For small $d$, even with a very large number of TLFs, the dynamics of decoherence is very far from Gaussian regime.

We remind the reader that Gaussian regime does not necessarily imply Gaussian form of the Ramsey or echo decay, but describes the situation when the decoherence dynamics is fully described by the first two moments of the random noise $B(t)$ in Eq.~\ref{eq:Ham}, the mean $\langle B(t)\rangle$ and the correlation function $\langle B(t) B(t+s)\rangle$.

(ii) We have demonstrated, using direct analytical calculation, see Sec.~\ref{sec:three} and Appendix, that the quantity $Z$ that controls the form of the Ramsey decay, is not distributed normally even for large $n$, unless $d$ is also large. In the situation when $d\ll 1$, the quantity $Z$ has the distribution density $P_Z(z)\propto z^{d-1}$, and the normal distribution is recovered only for $d\gg 1$.

This calculation is illustrated and confirmed by the numerical simulations, and the nature of the transition from the regime of sparse bath to the dense bath regime is explained in detail.

(iii) We have demonstrated, using numerical simulations, that the noise spectral density, or, equivalently, the correlation function of the noise $B(t)$ produced by the bath of TLFs, shows very little sample-to-sample variability for different sets of parameters $\mathcal{B}$ describing individual TLFs comprising the bath. For both dense and sparse baths, the noise spectra $S(\mathcal{B};\omega)$ demonstrate very similar $1/f$ form (with cutoffs at $\gamma_m$ and $\gamma_M$), with little variation in the magnitude or in shape, for different sets $\mathcal{B}$.

(iv) In contrast, the Ramsey decay curves in the case of sparse baths show significant sample-to-sample variability. The same level of large variability is seen in the echo decay curves produced using different bath parameter sets. However, dense baths do not show such a variability, with different samples producing very similar Ramsey and echo decay curves.

(v) Therefore, we demonstrate, using both analytical and numerical tools, that the situation of sparse baths is not described by the conventional theory, which predicts direct relation between the noise power spectrum $S(\omega)\propto 1/\omega$ and the dynamics of decoherence, i.e.\ the form of the Ramsey decay,  echo decay, and the coherence decay for various dynamical decoupling sequences.

(vi) We have demonstrated, using both analytical and numerical tools, that the Ramsey decay and the echo decay in the case of sparse baths are almost completely controlled by the parameters of only one or two TLFs, called here exceptional TLFs. We identified these fluctuators for both Ramsey and echo decoherence. 

We have directly demonstrated the role of these special TLFs by removing them from the bath in our simulations, and comparing the Ramsey and the echo decay curves before and after removal. We have explicitly shown that removal of the exceptional TLFs lead to significant slowdown of the Ramsey and the echo decay. We have shown that this happens for both individual bath parameter samples, and for ensembles of sparse baths.

We emphasize here that the strong effect of the exceptional TLFs on decoherence is not caused by their particularly strong coupling to the qubit. It is caused by the special values of their switching rates, and therefore represents an effect caused by the intrinsic dynamics of the bath.

(vii) We have shown that the removal of the exceptional TLFs can greatly improve the coherence properties of the qubit ensembles. To characterize this improvements in the case of Ramsey decay, we used the decay time $T_2^*$ and the qubit fidelity (defined as the value of the Ramsey decay function $F(t_0)$ at a certain small value of time $t_0$). We have shown that the fraction of high-quality qubits, where these metrics exceed certain threscholds, increases by a factor of 3--4, and sometimes even by a factor of 18, upon removal of only two exceptional TLFs from the baths comprising a total of 10--20 fluctuators.

Similar study was performed for the echo decay, and similar conclusions were reached. The echo decay time $T_2$ and the echo fidelity (defined in analogy with the Ramsey fidelity) were used as the metrics. The simulations have shown that the fraction of high-quality qubits increases by a factor of 3--4, and sometimes even by a factor of 18, upon removal of only two exceptional TLFs from the baths comprising a total of 10--20 fluctuators.

We note that it has been demonstrated in experiments \cite{Meyer2023electrical,Lisenfeld2023enhancing,RyuKangDevitalizeDefects22} that individual defects can be deactivated in the already fabricated devices by adjusting the working point parameters. If the exceptional TLFs, identified in this work, can be deactivated or avoided at the fabrication stage, that would greatly help in creation of large-scale qubit registers with uniformly high quality of the qubits. We do not discuss, how realistic this possibility is; such a verdict is to be made by experimentalists, material scientists and engineers. We point out that our theoretical investigation demonstrated great potential benefits of removal of only 1--2 exceptional TLFs from a bath made of many fluctuators.

(viii) We have demonstrated that our findings hold in the situation where the coupling strengths of individual TLFs have finite dispersion. We have considered the normally distributed coupling strengths, and saw the same effects even when the standard deviation $\sigma_v$ was equal to the mean $\bar{v}$ coupling strength. Thus, we have confirmed that the unusual effects associated with the sparse baths are not artifacts of the particular model, but represent real effects, which can be observed in real devices.

(ix) We have briefly studied more complex modes of control, investigating dephasing of a qubit coupled to a sparse bath and subjected to a train of closely spaced decoupling pulses (Carr-Purcell sequence). We have demonstrated that, contrary to possible expectations, a large number of pulses leads to rather modest improvements in the decoherence rate. However, it greatly reduces the sample-to-sample variability in the signal decay curves, which seem to acquire Gaussian shape. We provided an analytical explanation for this effect. More detailed investigation devoted to DD of qubits coupled to sparse baths of TLFs is deferred to subsequent work.

Therefore, our work has provided several important insights: firstly, into the nature of the Gaussian and non-Gaussian regimes in the baths made of many TLFs and exhibiting $1/f$ type of noise; secondly, into the role played by individual defects in the baths comprised of many fluctuators; and thirdly, into the possible ways of improving the quality of the solid-state qubits. Moreover, our work offers a possible path towards bridging conventional theories, ascribing decoherence of semi- and superconducting qubits to the baths made of many defects, and the recently suggested theories  \cite{Schloer2019,Ahn2021,Connors2020} emphasizing the role of individual defects.


\section*{Acknowledgements}
We thank L.~M.~K.~Vandersypen, M.~Rimbach-Russ, L.~DiCarlo, M.~Veldhorst, M.~Meyer, B.~P.~Wuetz, B.~Undseth and O.~Pietx-Casas for helpful discussions on quantum dot spin qubits and superconducting qubits. 
This work is part of the research programme NWO QuTech Physics Funding (QTECH, programme 172) with project number 16QTECH02, which is (partly) financed by the Dutch Research Council (NWO). The work was partially supported by the Kavli Institute of Nanoscience Delft.


\bibliography{refsMM-VD-stlf}

\appendix

\section{Ramsey decay of a qubit coupled to a sparse bath}\label{AppendixCohFidelity}

In order to gain analytical insight into the unusual features of the Ramsey decay of a qubit coupled to a sparse bath, we consider the situation when all TLFs are motionally narrowed, with $\gamma_k\gtrsim \bar{v}$ for all $k=1,\dots n$. In that case, the Ramsey decay has the form (see Eqs.~\ref{eq:RamseyHahnProduct} and \ref{eq:RamseySingleTLF}) 
\begin{equation}
F(t) \approx \exp{[-Z {\bar{v}}^2 t/2]},
\end{equation}
where
\begin{equation}
\quad Z=\sum_{k=1}^n \gamma_k^{-1},
\end{equation}
such that we should analyze the statistics of the quantity $Z$, and the region of most interest is the region of small $Z$, where its contribution to the function $F(t)$ is the largest. We focus primarily on the situation of a sparse bath, where the density of TLFs is small, $d\ll 1$.

For convenience, let us denote $x_k=\gamma_k^{-1}$. Since we assume that $\gamma_k$ are independent identically distributed (i.i.d.) variables, so are the quantities $x_k$, and each $x_k$ has exactly the same log-uniform p.d.f.\ $P(x_k)$ as the rate $\gamma_k$, namely
\begin{equation}
\label{eq:pdfx}
P(x) = \frac{1}{w\, x},\qquad x\in[x_m,x_M],
\end{equation}
with the width $w=\ln{[\gamma_M/\gamma_m]}$ and sharp cutoffs at $x_M=1/\gamma_m$ and $x_m=1/\gamma_M$; note that the lower cutoff of $x_k$ is the upper cutoff of $\gamma_k$ and vice versa, such that $w=\ln{[x_M/x_m]}$.

\begin{figure*}[t!]
\includegraphics[width=\linewidth]{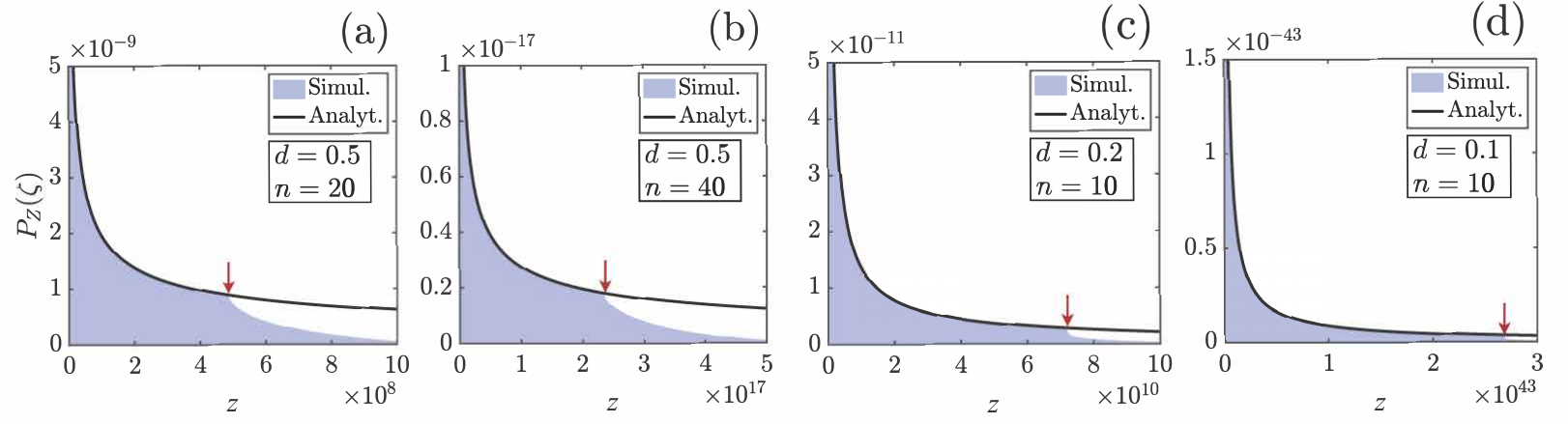}
\caption{Histograms (blue) estimating the p.d.f.\ of $P_Z(z)$ for different values of the bath density $d$ and the number $n$ of TLFs. Solid black curves show the analytically obtained dependence $P_Z(z)\propto z^{d-1}$. The p.d.f.'s are evaluated using $10^8$ samples of $Z$.}
\label{fig:FigApp2} 
\end{figure*}

In order to find the p.d.f.\ $P_Z(z)$, we use the slightly modified Markov method as presented in Ref.~\cite{ChandrasekharRandomProc}. Since
\begin{equation}
Z=\sum_{k=1}^n x_k,
\end{equation}
its p.d.f.\ can be written as
\begin{eqnarray}
P_Z(z) &=& \int \delta(Z-z) \prod_{k=1}^n P(x_k) dx_k \\ \nonumber
&=& \int \delta\left(\sum_{k=1}^n x_k - z\right) \prod_{k=1}^n P(x_k) dx_k.
\end{eqnarray}
Using the representation 
\begin{equation}
\delta(Z-z) = \frac{1}{2\pi} \int_{-\infty}^{+\infty} \exp{[i\lambda (Z-z)]}\, d\lambda,
\end{equation}
we find that 
\begin{equation}
P_Z(z) = \frac{1}{2\pi} \int_{-\infty}^{+\infty} \exp{[-i\lambda z]}\,P_n(\lambda)\, d\lambda,
\end{equation}
where
\begin{eqnarray}
\nonumber
P_n(\lambda) &=& \int \prod_{k=1}^n \mathrm{e}^{i\lambda x_k}\, P(x_k) dx_k  
= \left[\int \mathrm{e}^{i\lambda x}\, P(x) dx \right]^n \\
&=&
\left[\frac{1}{w}\,\int_{x_m}^{x_M} \frac{\mathrm{e}^{i\lambda x}}{x} dx \right]^n,
\label{eq:pn}
\end{eqnarray}
since all $x_k$ are i.i.d. variables with the p.d.f.\ given by Eq.~\ref{eq:pdfx}

We are interested in macroscopically (or at least mesoscopically) large baths, with $n\gg 1$, but finite (and small) density $d$. Formally, this means that we consider the limit $n\to\infty$ and $w\to\infty$ with fixed $d=n/w$. In this limit, we must be careful with the integral (\ref{eq:pn}) because of the associated singularities, which determine its behavior, especially in non-Gaussian regimes, beyond the region described by the central limit theorem \cite{KlauderAnderson62}, see detailed discussion in Ref.~\onlinecite{HahnDobrovitski21}.

Therefore, we are employing the method suggested in Ref.~\onlinecite{KlauderAnderson62}, and re-write the integral as
\begin{eqnarray}
P_n(\lambda) &=& \left[1 - \int \left(1-\mathrm{e}^{i\lambda x}\right) P(x) dx \right]^n \\ \nonumber
&=& \left[1-\frac{1}{w}\,\int_{x_m}^{x_M} \frac{1-\mathrm{e}^{i\lambda x}}{x} dx \right]^n   
\end{eqnarray}
using the fact that $\int P(x) dx = 1$. In the limit $n\to\infty$ and $w\to\infty$, we can re-write the latter integral as 
\begin{equation}
\label{eq:ylambda}
P_n(\lambda) = \mathrm{e}^{- d\, Y(\lambda)},\quad 
Y(\lambda) = \int\limits_{x_m}^{x_M} \frac{1-\mathrm{e}^{i\lambda x}}{x} dx. 
\end{equation}
using the fact that $1/w=d/n$ and the limit $(1-a/n)^n\to\exp{(-a)}$ at $n\to\infty$.

Now we need to address the issue of the limits: the physically meaningful situation is when $x_m\to 0$ and $x_M\to\infty$. The lower limit, $x_m\to 0$, can be eliminated in a rather straightforward manner, by expressing the integral $Y(\lambda)$ via the exponential integral function $\mathrm{Ein}(z)$ \cite{AbramStegun,DLMF}, which is analytic on the whole complex plane and vanishes at $z\to 0$. Therefore, we replace $x_m$ by 0, obtaining 
\begin{equation}
\label{eq:xM}
Y(\lambda) = \int_0^{x_M} \frac{1-\mathrm{e}^{i\lambda x}}{x} dx. 
\end{equation}
The upper limit is to be kept as is; there are two reasons for that. The first reason is quite formal, the function $Y(\lambda)$ diverges at $x_M\to\infty$. The second reason is more subtle, being associated with the appearance of the non-Gaussian regime, and will be discussed and clarified later.

We consider the limit of $x_M\to\infty$, where asymptotic behavior of $Y(\lambda)=Y_1(\lambda) + i Y_2(\lambda)$ is known \cite{AbramStegun,DLMF}:
\begin{eqnarray}
Y_1(\lambda) &=& \ln{(\gamma_E x_M |\lambda|)}+O(\mathrm{e}^{-x_M |\lambda|}),\\ \nonumber
Y_2(\lambda) &=& -\,\frac{\pi}{2}\,\mathrm{sgn} \lambda +O(\mathrm{e}^{-x_M |\lambda|}),
\end{eqnarray}
where $\gamma_E$ is the Euler-Mascheroni constant and $\mathrm{sgn}\lambda$ is the sign function. Omitting the exponentially small terms in the expressions above, we have
\begin{eqnarray}
\mathrm{e}^{Y(\lambda)} &=& -i\gamma_E x_M \lambda,\\ \nonumber
\mathrm{e}^{-d\,Y(\lambda)} &=& (-i\gamma_E x_M \lambda)^{-d} = (\gamma_E x_M |\lambda|)^{-d}\,\mathrm{e}^{i\,\frac{\pi d}{2}\,\mathrm{sgn}\lambda}.
\end{eqnarray}
Thus, we obtain 
\begin{eqnarray}
\label{eq:sing}
P_Z(z) &=& \frac{1}{2\pi} \int\limits_{-\infty}^{+\infty} \mathrm{e}^{-i\lambda z} \mathrm{e}^{-d\,Y(\lambda)}\, d\lambda \\ \nonumber
&=& \frac{1}{2\pi} \int\limits_{-\infty}^{+\infty} \mathrm{e}^{-i\lambda z}
(\gamma_E x_M |\lambda|)^{-d}\,\mathrm{e}^{i\,\frac{\pi d}{2}\mathrm{sgn}\lambda}\, d\lambda.
\end{eqnarray}
We immediately see that the integral has the singularity $\lambda^{-d}$, which is a weak integrable singularity for $d\ll 1$, but becomes strong for $d\gg 1$.

\begin{figure*}[tph!]
\includegraphics[width=1\linewidth]{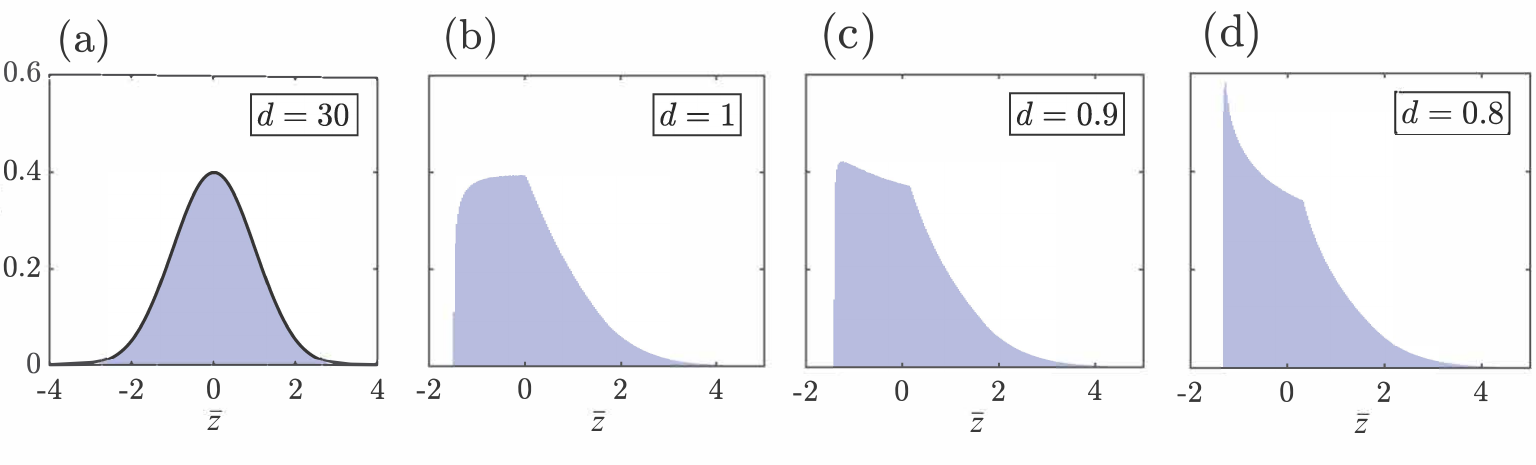}
\caption{Qualitative changes in the form of the p.d.f.\ $P_Z(z)$, taking place for increasing density $d$. To make the qualitative comparison more clear, the values of $Z$ were rescaled, such the figures show the p.d.f.'s of the rescaled 
quantity $\bar{Z}=(Z - \mu_Z)/\sigma_Z$, where $\mu_Z$ and $\sigma_Z$ are the mean and the standard deviation of $Z$, respectively. Each probability density is estimated using $10^8$ samples of $Z$. The number of TLFs, $n=20$, remains the same while $d$ decreases from $d=30$ in panel (a) to $d=0.8$ in panel (d).  For large density, $d=30$, the function $P_{\bar{Z}}(\bar{z})$ practically coincides with the normal distribution (solid black curve).}
\label{fig:FigApp1}
\end{figure*}

For dense baths with $d\gg 1$, behavior of the integral (\ref{eq:sing}) is controlled by this singularity, i.e.\ by the region where $\lambda$ is small. This is precisely the region where we replaced the finite quantity $x_m=1/\gamma_M$ by zero. Now we see that such a replacement is allowed only for $d<1$, where $\lambda^{-d}$ is integrable, and is prohibited otherwise, especially for large values of $d$. In fact, it is not difficult to repeat the calculations above keeping $x_m$ finite and consider the asymptotics $d\to\infty$, practically reproducing the classical calculations presented in  Ref.~\onlinecite{ChandrasekharRandomProc}. The answer is precisely what one would anticipate, the normal distribution $P_Z(z)$. Thus, the Gaussian part of the function $P_Z(z)$ is controlled by the function $P_n(\lambda)$ in the region of small $\lambda$, and this is what determines the form of $P_Z(z)$ for $d\gg 1$.

However, as we already mentioned, this part is negligible for small $d$, and in that case the function $P_Z(z)$ is controlled by the regions where $\lambda$ is large. Continuing the last line of Eq.~\ref{eq:sing}, and considering the regions $\lambda>0$ and $\lambda<0$ separately, we obtain 
\begin{equation}
P_Z(z) = \frac{1}{\pi} \int\limits_{0}^{+\infty} (\gamma_E x_M \lambda)^{-d}\, \cos{\left(\lambda z-\frac{\pi d}{2}\right)}\,d\lambda,
\end{equation}
and, with the change of variable $\nu=\lambda z$, we find
\begin{equation}
P_Z(z) = \frac{1}{\pi}\,\frac{z^{d-1}}{(\gamma_E x_M)^d}\, 
\left(G_c \cos{\frac{\pi d}{2}} + G_s \sin{\frac{\pi d}{2}}\right),
\end{equation}
where \cite{janson}
\begin{eqnarray}
G_c &=& \int_0^\infty \frac{\cos{\nu}}{\nu^d}\,d\nu = 
\cos{\frac{\pi(1-d)}{2}}\,\Gamma(1-d)\\ \nonumber
G_s &=& \int_0^\infty \frac{\sin{\nu}}{\nu^d}\,d\nu = 
\sin{\frac{\pi(1-d)}{2}}\,\Gamma(1-d).
\end{eqnarray}
where $\Gamma(\cdot)$ is the Gamma function, and we obtain the answer we were looking for:
\begin{equation}
P_Z(z) = z^{d-1}\, \frac{\Gamma(1-d)}{\pi}\,\frac{\cos{\frac{\pi (2d-1)}{2}}}{(\gamma_E x_M)^d},
\end{equation}
or, simply,
\begin{equation}
P_Z(z) \propto z^{-(1-d)}.
\end{equation}
The essence of the calculations performed in this Appendix can be extracted, producing the simplified qualitative argument presented in Sec.~\ref{sec:three} of the main text.

The result $P_Z(z) \propto z^{-(1-d)}$ is confirmed in numerical calculations, where $n$ values of $\gamma_k$ were sampled from the distribution (\ref{eq:pdfx}) and the value $Z=\sum_k \gamma_k^{-1}$ was calculated. Performing this sampling $10^8$ times, the results shown in Fig.~\ref{fig:FigApp2} were obtained, demonstrating very good agreement with the analytical calculations even at rather large density $d=0.5$.

Note that the analytical calculations above have been performed in the limit $n\gg 1$, and assuming that $x_m\ll 1$ and $x_M\gg 1$. In numerical simulations, with finite values of $n$, $x_m$ and $x_M$, there are no values of $Z$ in the region $Z<n\cdot x_m$. Besides, the p.d.f.\ $P_Z(z)$ differs from the analytical prediction for $Z>(n-1)x_m + x_M$: this region corresponds, in essence, to a bath made of $n-1$ TLFs, with the remaining TLF having $x=x_M$. Formally, Eq.~\ref{eq:xM} needs to be corrected in that region of $Z$. The boundary of the region $Z>(n-1)x_m + x_M$ is marked by a small red arrow in each panel of Fig.~\ref{fig:FigApp2}, and a cusp in $P_Z(z)$ is clearly seen there.

These edge effects of finite $x_m=\gamma_M^{-1}$ and $x_M=\gamma_m^{-1}$ are irrelevant in the limit considered here. However, they play an important role as the bath becomes denser, i.e.\ with increasing $d$.
We pointed out above that the integral $P_Z(z)$ for $d\gg 1$ is dominated by the region where $\lambda$ is small, such that the finite value of $x_m$ becomes crucial. In that case, as $d$ increases, the region $Z>n\,x_m$ (where the analytical calculations above belong) starts to overlap strongly with the region $Z>(n-1)x_m + x_M$, and with other similar regions, namely $(n-2)x_m + 2 x_M > Z > (n-1)x_m + x_M$ (where the next cusp in $P_Z(z)$ occurs), etc. As all these regions overlap more and more, the shape of $P_Z(z)$ changes, and finally, for $d\gg 1$, the function $P_Z(z)$ acquires Gaussian form, corresponding to the limit expected from the central limit theorem. This is illustrated in Fig.~\ref{fig:FigApp1}. Note that, as $d$ increases, the characteristic values of the quantity $Z$ change significantly; in order to make the comparison more clear, Fig.~\ref{fig:FigApp1} shows the estimated p.d.f.'s of the rescaled quantity, $\bar{Z}=(Z - \mu_Z)/\sigma_Z$, where $\mu_Z$ and $\sigma_Z$ are the mean and the standard deviation of $Z$. 

Finally, we quickly discuss the region of the actual bath parameters, where our analytical calculations and arguments are applicable. The approximation $f_k\approx\exp{(-{\bar{v}}^2 t\gamma_k^{-1}/2)}$ for the Ramsey decay under the action of the $k$-th TLF is valid when $t\gtrsim \gamma_k^{-1}$ and $v\lesssim\gamma_k$. Since the values of $\gamma_k$ are distributed log-uniformly, roughly half of the TLFs have $\gamma<\gamma_C$, where $\gamma_C=\sqrt{\gamma_m\cdot\gamma_M}$. Thus, roughly half of the TLFs are in the regime of motional narrowing when time $t$ becomes sufficiently large, $t\gtrsim t_C=\gamma_C^{-1}$. Before the time moment $t_C$ is reached, the Ramsey decay has quasi-static character, with $f_k\approx\exp{(-{\bar{v}}^2 t^2/2)}$. For our calculations to be valid, most of the Ramsey decay should happen in the regime of motional narrowing, such that the initial quasi-static decay should be negligibly small, i.e.\ 
\begin{equation}
\prod\limits_{k=1}^n \exp{(-{\bar{v}}^2 t_C^2/2)} = \exp{(-n {\bar{v}}^2 t_C^2/2)}\approx 1,
\end{equation}
which leads to the requirement 
\begin{equation}
\label{eq:req}
{\bar{v}}\ll t_C^{-1}/\sqrt{n} = \sqrt{\gamma_m\gamma_M/n}.
\end{equation}
This requirement is easily satisfied, since our analysis of the Ramsey decay is performed for the baths where $\gamma_M=\gamma_m\mathrm{e}^{n/d}$ with $d\lesssim 1$ or even $d\ll 1$. In that regime, in order to satisfy the requirement (\ref{eq:req}), we need only
\begin{equation}
{\bar{v}}\ll \gamma_m\cdot \mathrm{e}^{n/(2d)}/\sqrt{n},
\end{equation}
while $\mathrm{e}^{n/(2d)}/\sqrt{n}$ is a large quantity at $n\sim 10$ and $d\lesssim 1$.

\end{document}